\documentclass[fleqn,usenatbib]{mnras}
\usepackage{mathptmx}

\usepackage[T1]{fontenc}

\DeclareRobustCommand{\VAN}[3]{#2}
\let\VANthebibliography\thebibliography
\def\thebibliography{\DeclareRobustCommand{\VAN}[3]{##3}\VANthebibliography}

\usepackage{graphicx}	
\usepackage{amsmath}	
\usepackage{natbib}

\title[A MeerKAT Spectropolarimetric Study of Pictor A]{A Spectropolarimetric Study of Pictor A Radio Galaxy with MeerKAT}

\author[Andati et al.]{Lexy A. L. Andati$^{1}$\thanks{E-mail: andatilexy@gmail.com},
Lerato M. Baidoo$^{1,2}$,
Athanaseus J. T. Ramaila$^{1,3}$,
Oleg M. Smirnov$^{1,3}$,
\newauthor
Sphesihle Makhathini$^{4,1}$,
and Richard A. Perley$^{5,1}$ \\
$^{1}$Centre for Radio Astronomy Techniques and Technologies, Department of Physics and Electronics, Rhodes University, PO Box 94, Makhanda 6140, South Africa\\
$^{2}$Dunlap Institute for Astronomy and Astrophysics, University of Toronto, Toronto, ON M5S 3H4, Canada\\
$^{3}$South African Radio Astronomy Observatory, 2 Fir Street, Black River Park, Observatory, 7925, South Africa\\
$^{4}$School of Physics, University of the Witwatersrand, 1 Jan Smuts Avenue, Johannesburg 2000, South Africa\\
$^{5}$National Radio Astronomy Observatory, Socorro, NM 87801, USA
}

\date{Accepted XXX. Received YYY; in original form ZZZ}

\pubyear{2023}

\begin{document}
\label{firstpage}
\pagerange{\pageref{firstpage}--\pageref{lastpage}}
\maketitle

\begin{abstract}
We present the results of a polarimetric study from our new high-sensitivity L-band (0.8--1.7 GHz) observation of Pictor A with the MeerKAT radio telescope. We confirm the presence of the radio jet extending from the nucleus to the western hotspot of this source. Additionally, we show the radio emission expected to be coincident with previously observed X-ray emission in the radio lobes, confirming that the emission mechanism is of inverse Compton origin, as suggested by a previous study. Our spectropolarimetric analysis using the RM-Synthesis technique reveals a relatively uniform mean RM distribution across the lobes of Pictor A, with most lines-of-sight exhibiting single-peaked Faraday spectra. However, a number of the lines-of-sight exhibit single peaked spectra with a wide base or multiple peaks, suggesting the presence of multiple Faraday components or a Faraday thick structure along Pictor A's lines-of-sight. We also confirm the asymmetry in RM variability and depolarization between the two lobes of this source which were reported in a previous study.
\end{abstract}

\begin{keywords}
Polarization -- galaxies:magnetic fields -- galaxies: interactions -- galaxies: individual Pictor A
\end{keywords}



\section{Introduction}
Pictor A is the fifth-brightest discrete radio source in the Southern sky. It has a radio power of $3.79 \times 10^{26}$ W Hz$^{-1}$ \citep{Robertson1973} and is a FR-II galaxy \citep{Fanaroff1974}, with structure consisting of two radio lobes, hotspots at the edge of each lobe, a core at its centre, and radio jets. Its host galaxy is centred at right ascension (RA, J2000) 5$^{\textrm{h}}$19$^{\textrm{m}}$50$^{\textrm{s}}$ and declination (Dec) -45$^{\circ} $46\arcmin $44$\arcsec, and is identified as a broad-line galaxy \citep{Halpern1994,Lewis2010}, with disc-like morphology \citep{Inskip2010} of either Sa \citep{Lauberts1982} or S0 \citep{Loveday1982} type. Pictor A is at a redshift of $0.035$ \citep{schmidt1965}, such that 1\arcsec angular scale corresponds to 0.697 kpc in linear scale.\footnote{Assuming a $\Lambda$CDM cosmological model, with $H_{\textrm{0}} = 70$ km s$^{-1}$ $\Omega_{\textrm{m}} = 0.3$ and $\Omega_{\Lambda}$ = 0.7.} This radio source has an angular size of $\sim$8 arcmin in the East-West (EW) direction, and $\sim$4 arcmin in the North-South (NS) direction, which corresponds to $\sim$340 kpc EW and $\sim$170 kpc NS in linear size, respectively \citep[e.g. see fig. 3 of][]{perley1997}.

The radio lobes of Pictor A are known for their remarkably round shape -- with ellipticity > 0.9 \citep{perley1997}, which is unusual for FR-II galaxies. The lobes of FR-II galaxies are usually found to be elongated along a specific direction (e.g. the direction of jet propagation). Thus, \citet{perley1997} supposes that the lobes of Pictor A may be relaxed in a uniform environment -- hence its uniform expansion in all directions. Moreover, these lobes showed no edge brightening, except for filamentary structures particularly in the western lobe consisting of enhanced surface brightness of 0.5 mJy arcsec$^{-2}$. The presence of such filamentary structure is common in radio galaxies; for example, they are seen in the lobes of Cygnus A \citep{perley1984}, Fornax A \citep{Fomalont1989,Anderson2018}, 4C 12.03, CGCG 044-046 and CGCG 021-063 \citep{Fanaroff2021}. 
  
The X-ray emission spectra of Pictor A's lobes are best modelled using a power-law model with a photon index of 1.57$\pm$0.04 \citep{Hardcastle2016}, indicating that the dominant source of emission is the inverse Compton scattering of CMB photons off the relativistic electron population. Fitting a two-component model (power-law and thermal) to the region absent of radio emission, the authors found a `soft thermal emission' component of temperature 0.33$\pm$0.07 keV (also noted by \citealt{zirbel1997}) and a rather steep photon index of 2.07. The results were inconclusive, with no compelling evidence for including a thermal component.

The western hotspot (WHS hereon) is one of the brightest radio and X-ray hotspots known \citep[][]{Wilson2001,Hardcastle2004, Tingay2008}. The radio hotspot is identified to be coincident with a compact, very bright (magnitude of 19.5) and highly polarized ($>$30 per cent) source of optical emission \citep{Roeser1987}, and with an extremely bright X-ray hotspot \citep{Wilson2001}. In the X-ray, \citet{Hardcastle2016} determined this hotspot to be variable on scales of months to a year. The eastern lobe has two hotspots \citep{Prestage1985,perley1997}.

The western radio jet was first observed by \citet{perley1997}, with no evidence of a counter-jet (the eastern jet). In the X-ray, the western jet was observed first by \citet{Wilson2001}, and the counter-jet by \citet{hardcastle2005, Hardcastle2016}, while in the optical, jet knots coincident with the known radio and X-ray jet were imaged by \citep{gentry2015}. The western jet (in the X-ray) extends from the core out to a distance of 250\arcsec \citep[174 kpc,][]{Hardcastle2016}. The jet emission responsible for the observed X-rays is purely synchrotron radiation \citep{Hardcastle2016}. At 34 and 49 kpc from the core, the X-ray jet is observed to be variable \citep{Marshall2010,Hardcastle2016}.

The radio emission across the lobes of Pictor A is linearly polarized and is undergoing Faraday rotation. \citet{perley1997} presents this source's most detailed polarimetric study using the Very Large Array (VLA) telescope at 90, 20, 6, 3.6 and 2 cm and at a few arcseconds resolution. \citet{perley1997} found significant polarization across the source, ranging between 30 to 60 per cent along the lobe edges and between 10 to 20 per cent within the central regions of the lobes. The authors derived the mean rotation measures (\textrm{RM}) of both lobes to be $43.5\pm1.4$ rad m$^{-2}$, which is consistent with previous studies. The challenge faced by the VLA observations was that the source was too far South. Therefore, Pictor A never exceeded an elevation of 10$^\circ$ due to the telescope's latitude. This resulted in a significantly limited observation time ($\sim$4 hrs), hence reduced sensitivity and low fidelity of the output images. Furthermore, the foreshortening of the N-S baselines due to the low source elevation resulted in an elongated synthesised beam. 

In this work, we present a study of Pictor A using high-sensitivity L-band (0.8--1.7 GHz) data from the MeerKAT telescope. The location of this telescope made it a suitable instrument for the observation of this source. In Sec.~\ref{sec:total-intensity-features}, we present the total intensity features of Pictor A from our observations, including the radio jet and the `missing' accompanying radio emission expected to be co-spatial with previously observed IC/CMB emission \citep{Hardcastle2016}. Sec.~\ref{sec:polarimetry-results} presents a new spectropolarimetric analysis of Pictor A using newer diagnostic methods: RM-synthesis and $\mathit{QU}$-fitting. This is followed by a summary and conclusion.

\section{Observations and Data Reduction}
  Pictor A was observed in December 2019 using the MeerKAT L-band system in full polarization 4K mode, whereby each channel width was 208.984 kHz. The specific details of the observation are summarised in Table ~\ref{tab:obs-info}. 
    \begin{table}
      \centering
      \caption{Observational setup}
      \label{tab:obs-info}
      \begin{tabular}{cc}
          \hline
          \multicolumn{2}{|c|}{Observation Details for Pictor A}\\
          \hline
          Dates (UTC) & from: 2019 Dec 18 16:46 \\
          & to: 2019 Dec 19 03:48 \\
          Antennas & 62 \\
          Band & L \\
          Frequencies & 856 MHz -- 1.7 GHz \\
          Bandwidth & 856 MHz \\
          Channels & 4096 \\
          Time on target & $\sim$ 9 Hrs \\
          Scan duration & $\sim$ 15 min \\
          Integration time & 8 sec \\
          Bandpass calibrator & J0408-6545 \\
          Gain calibrator & J0538-4405 \\
          Polarization Angle calibrator & J0521+1638 \\
          \hline
      \end{tabular}
  \end{table}

\subsection{Calibration}
The standard calibration procedures were undertaken using the \texttt{CARACal} pipeline \citep{jozsa2020} to obtain our polarimetry-ready data.\footnote{\url{https://caracal.readthedocs.io/en/latest/}} The procedures consisted of first-generation calibration (1GC) followed by the second-generation calibration (2GC), \citep[see][for a description of 1GC and 2GC]{smirnov2011b}. The 1GC comprises mostly of cross-calibration, whereby a well-known calibrator is used to correct the data of the target source, while the 2GC consists mostly of self-calibration (selfcal). The following calibration strategy was used:
\begin{enumerate}
  \item A static mask was applied to the already known RFI-affected regions in the MeerKAT L-band within the calibrator fields. Additional flagging was then performed to remove any remaining RFI (1200 $\sim$ 1230 MHz, 1280 $\sim$ 1300 MHz, 1560 $\sim$ 1610 MHz) and the band edges (856 $\sim$ 880 MHz and 1700 $\sim$ 1800 MHz). Furthermore, antennas and scans whose signals seemed corrupted from visual inspection were also excised at this step.
  \item Delay, complex gain, bandpass calibration, and flux scaling solutions were derived using J0538-4405 and J0408-6545.
  \item For polarization calibration, J0521+1638 was used to calibrate the polarization position angle, while the bandpass calibrator (J0408-6545) was used for leakage corrections.
  \item The obtained solutions were first applied to all the calibrator sources, where they were inspected for validity. Once determined to be suitable, they were transferred to the target source, marking the end of 1GC.
  \item Our first image of the target source generated from these calibrated data was used as our model image for our initial selfcal.
\end{enumerate}

Self-calibration proceeded in two stages:

\begin{itemize}
    \item During the first stage, we performed two rounds and phase-and-delay selfcal, using Stokes $\mathit{I}$ models only. In particular, we used the \texttt{CubiCal} calibration software \citep{kenyon2018cubical}, selecting a `unislope' solver with a solution interval of 4096 channels in frequency (the full band) and a single integration in time. In effect, this solves for independent X and Y phase offsets and a common phase slope (i.e. delay term) per each solution interval.

    \item For the second stage, we performed phase-and-amplitude self-calibration using Stokes $\mathit{I, Q}$ and $\mathit{U}$ models. This used \texttt{CubiCal}'s G complex 2-by-2 solver with a solution interval of 64 channels in frequency and 8 integrations in time. No Stokes $\mathit{V}$ models were used.
\end{itemize}

However, Pictor A's extremely bright western hotspot introduced major artefacts into our images and, thus, posed a great challenge for our 2GC. As similarly noted by \citet{Hardcastle2016}, it required different calibration solutions from the rest of the source. We applied the differential gains \citep{smirnov2011} iteratively to improve our models for this hotspot to obtain better images. The specific steps undertaken for calibration and high dynamic range imaging will be highlighted in a different paper: Ramaila et al. (in prep). 

\begin{figure*}
    \centering
    \includegraphics[width=2\columnwidth]{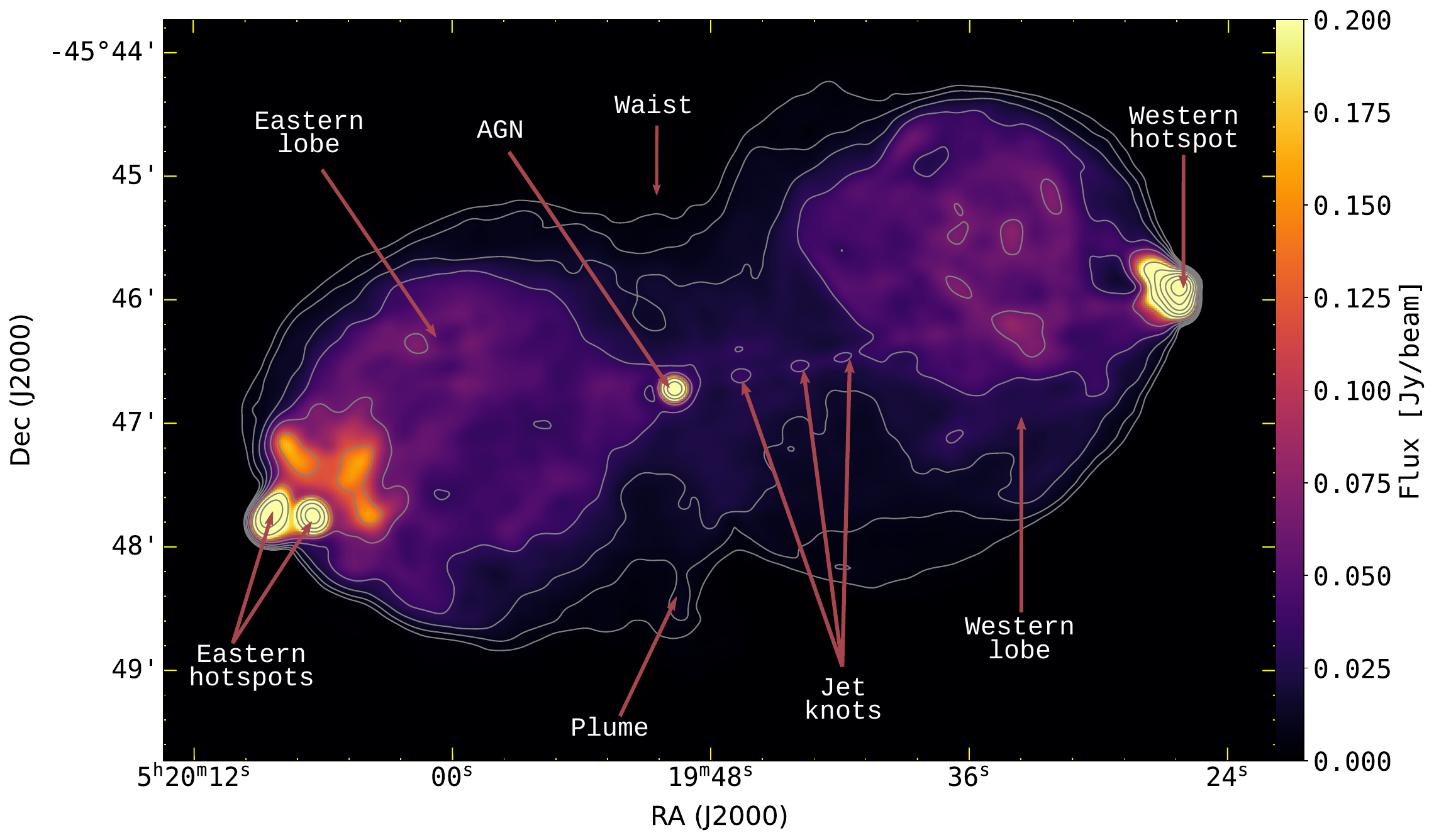}
    \caption{An MFS image of Pictor A from our new MeerKAT L-band observations centred at 1.28 GHz with a bandwidth of 856 MHz and $\sim$7.5\arcsec resolution. The contour levels shown start from 4 mJy beam$^{-1}$ and increase by a factor of 2. The off-source noise is 22 $\mu$Jy beam$^{-1}$. The source emission shows a single hotspot at the extreme end of the western lobe and two hotspots at the extremity of the eastern lobe. The jet radio knots on the west side of the source are seen with a clear path from the Active Galactic Nucleus (AGN) to the hotspot.}
    \label{fig:pica}
\end{figure*}
During imaging, we averaged our data in frequency, ensuring that our channel widths were small enough to avoid bandwidth depolarization and large enough to maintain a good signal-to-noise ratio, thus improving sensitivity. We used a maximum channel width, $\delta\nu $, of $\sim$ 10.7 MHz, resulting in 80 sub-band images. We did not apply primary beam corrections to the whole field overall as it seemed unnecessary for these data since our interest was the central source (Pictor A) whose angular size is $\sim$8 arcmin -- well within the MeerKAT main beam of $1.2$ degrees. However, differential gains were applied to the WHS.

All Imaging was done with the \texttt{wsclean} imaging and deconvolution software \citep{offringa_wsclean_2014} in multi-scale mode because of extended emission of this source. Upon inspection, multiple channels exhibiting abnormal spectral variations were identified and discarded. These channels coincided with the parts of our bandwidth that were heavily flagged because of GSM-induced radio frequency interference (RFI), perhaps leading to a reduced signal-to-noise. The remaining images were combined to form Stokes $\mathit{I}$, $\mathit{Q}$ and $\mathit{U}$ cubes which were all convolved to the same beam (11\arcsec by 10\arcsec) using the \texttt{spimple} \texttt{Python} package.\footnote{\url{https://github.com/landmanbester/spimple}} The resulting images were then used for polarimetric analysis of Pictor A.

We estimated an off-source noise of $21.85\mu$Jy beam$^{-1}$ for the $\mathit{I}$ multi-frequency synthesis (MFS) image, while the per-channel noise ranged between 95--278 $\mu$Jy beam$^{-1}$ in $\mathit{Q}$, and 41--233 $\mu$Jy beam$^{-1}$ in $\mathit{U}$. The noise close to the WHS increases by almost an order of magnitude for each image.

\subsection{Extracting Spectropolarimetric Information}
\label{sec:polarimetry}
To study the polarimetric composition of Pictor A, we applied the Rotation Measure synthesis (RM-synthesis) technique first introduced by \citet{Burn1966} and later refined by \citet{Brentjens2005}. RM-synthesis mitigates the $n\pi$ ambiguity problem that plagues the traditional linear fitting of the polarization angle against $\lambda^{2}$. In this context, erroneous values of \textrm{RM} and zeroth-wavelength position angle may be obtained due to polarization angle wrapping (or jumps) into multiple turns, hence the ambiguity. An in-depth discussion of RM-synthesis and its concepts can be found at \citet{Brentjens2005,heald2009a,ferriere}. Here, we briefly define some concepts related to our application of the technique.
Following \citet{Burn1966}, we define complex polarization as $P =  \mathit{Q} + i\mathit{U}$.

Polarization amplitude and angle can be derived from Stokes $\mathit{Q}$ and $\mathit{U}$ images respectively as follows:
\begin{align}
    |P| &= \sqrt{\mathit{Q}^2 + \mathit{U}^2} \text{  and },\label{eqn:pol-amp}\\
    \chi &= 0.5 \arctan \left(\mathit{\frac{U}{Q}}\right) \text{[rad]}.
\end{align}
As polarized emission is radiated from a source towards the observer, it encounters a variety of media that can perturb the intrinsic properties of the signal. In particular, when the emission passes through regions of magneto-ionized plasma, its polarization angle gets rotated due to the birefringent effect. This phenomenon is known as Faraday rotation. In the presence of Faraday rotation, the observed polarization angle, $\chi$, is given by:
\begin{equation}
    \label{eqn:chi_change}
    \chi = \chi_{0} + \mathrm{RM } \lambda^{2},
\end{equation}
where $\lambda$ is the observation wavelength, $\chi_{0}$ is the intrinsic polarization angle, and RM is known as the rotation measure. However, since multiple intervening magnetized media (both Faraday rotating and synchrotron emitting) could exist along a line-of-sight, \citet{Burn1966} introduced the notion of Faraday depth, $\phi$, given as: 
\begin{equation}
    \label{eqn:rm}
    \phi (L) = 0.81 \int_{\text{L}}^{\text{obs}} n_{e} \text{ \textbf{B} } \cdot d\mathbf{l} \text{ [rad m $^{-2}$] },
\end{equation}
where $n_{e}$ is the electron density in cm $^{-3}$, \textbf{B} is the magnetic field in $\mu$G, and $d\mathbf{l}$ is the infinitesimal path length in parsecs, and L is a specific point along the line-of-sight. This gives the RM contribution within a specific distance along the line-of-sight to an observer.

For a more general representation, contributions from the magnetised plasma and the multiple sources of Faraday rotation along the line-of-sight are taken into account by summing them up. Practically, we can only observe at $\lambda^2 > 0 $ at discretely sampled wavelengths. This means that the observed $P(\lambda^{2})$ is weighted by a sampling function $W(\lambda^2)$ -- whose values are only valid at the sampled wavelengths. Furthermore, \citeauthor{Brentjens2005} show that RM-synthesis operates within two Fourier-related spaces, namely the Faraday depth ($\phi$) space and lambda ($\lambda^{2}$). Taking the above limitations into consideration, the observed polarized intensity in $\lambda^2$ and $\phi$ is represented as follows:
\begin{align}
    \tilde{P}(\lambda^{2}) &= W(\lambda^{2}) P(\lambda^{2}) =  W(\lambda^{2})\int_{-\infty}^{+\infty} F(\phi) e^{2i\phi(\lambda^{2} - \lambda^{2}_{0})} d\phi,\\
    \tilde{F}(\phi) &= F(\phi) \star R(\phi)  = K \int_{-\infty}^{+\infty} \tilde{P}(\lambda^{2}) e^{-2i\phi(\lambda^{2} - \lambda^{2}_{0})} d\lambda^2 \label{eqn:fdf},
\end{align}
where $\tilde{P}$ and $\tilde{F}$ represent the observed quantities of $P$ and $F$, and $F(\phi)$ is the Faraday dispersion function (FDF henceforth): the polarized emission at each Faraday depth, and 
\begin{equation*}
    K = \left( \int_{\infty}^{\infty} W(\lambda^{2}) d\lambda^{2}\right)^{-1}.
\end{equation*}
From Eqn.~\ref{eqn:fdf}, the observed (dirty) Faraday spectra, $\tilde{F}$, results from the convolution of the true (clean/deconvolved) spectra $F(\phi)$ and $R(\phi)$ is the rotation measure transfer function (RMTF), a complex-valued function given by:
\begin{equation}
    \label{eqn:rmtf}
    R(\phi) = K \int_{-\infty}^{+\infty} W(\lambda^{2}) e^{-2i\phi(\lambda^{2} - \lambda^{2}_{0})} d\lambda^{2}.
\end{equation}
$\lambda^{2}$ is derotated to a reference $\lambda^{2}_{0}$ in Eqn.~\ref{eqn:rmtf} to mitigate difficulties in estimating the correct polarization angle at the peak of $|F(\phi)|$ due to the rapid rotation of the real and imaginary components of the RMTF.

The preceding discussion assumes that $Q$ and $U$ are independent of frequency except for the effects of Faraday rotation. In practice, there is an intrinsic spectral dependence \citep[Sec.3]{Brentjens2005}. In order to correct for this, we applied RM synthesis to the fractional quantities $q = Q/I$ and $u = U/I$ rather than to \textit{Q} and \textit{U}. We used the individual sub-band Stokes $\mathit{I}$ obtained from multiscale deconvolution to obtain these maps. We attempted spectral fitting on the Stokes $\mathit{I}$ image cube to obtain a Stokes $\mathit{I}$ model. However, using the model Stokes $\mathit{I}$ showed no significant improvements in the behaviour of the fractional polarization. Therefore, we used the originally obtained Stokes $\mathit{I}$ cubes.

The maximum Faraday depth, $|\phi_{\mathrm{max}}|$, we can probe is dependent on the channel width $\delta\lambda^{2}$ of our observations as follows:
\begin{equation}
    \label{eqn:phi_max}
    |\phi_{\mathrm{max}}| = \frac{\sqrt{3}}{\delta\lambda^{2}}.
\end{equation}
We mentioned in the previous section that our data were frequency averaged, yielding 80 sub-band images for the spectropolarimetric analysis. These frequency averaging bins were determined by deriving the maximum allowable channel width, $\delta\nu $, using the following equation \citep[a modification of Eqn. 4 of ][]{Sebokolodi2020}:
\begin{equation*}
    \delta \nu = \frac{\Delta \chi \nu}{2\lambda^{2}\mathrm{RM} },
\end{equation*}
where $\nu$ is the centre of the observational bandwidth, $\lambda$ is its corresponding wavelength, $\Delta \chi$ is the per-channel maximum allowable rotation of the plane of polarization in radians, and RM is the maximum rotation measure of the source. Pictor A's maximum rotation measure was shown by \citet{perley1997} as $\sim$ 100 rad m$^{-2}$. Therefore, using a condition that the maximum allowable rotation is 5 degrees and a centre frequency of 1.28 GHz, we derived $\delta \nu$ to be $\sim$ 10.2 MHz, which we adjusted to $\sim$ 10.7 MHz. By converting the chosen $\delta \nu$ to $\delta \lambda^{2}$, we then obtained a $\phi_{\text{max}}$ of $\sim$1981 rad m$^{-2}$.

\begin{figure}
    \centering
    \includegraphics[width=\columnwidth]{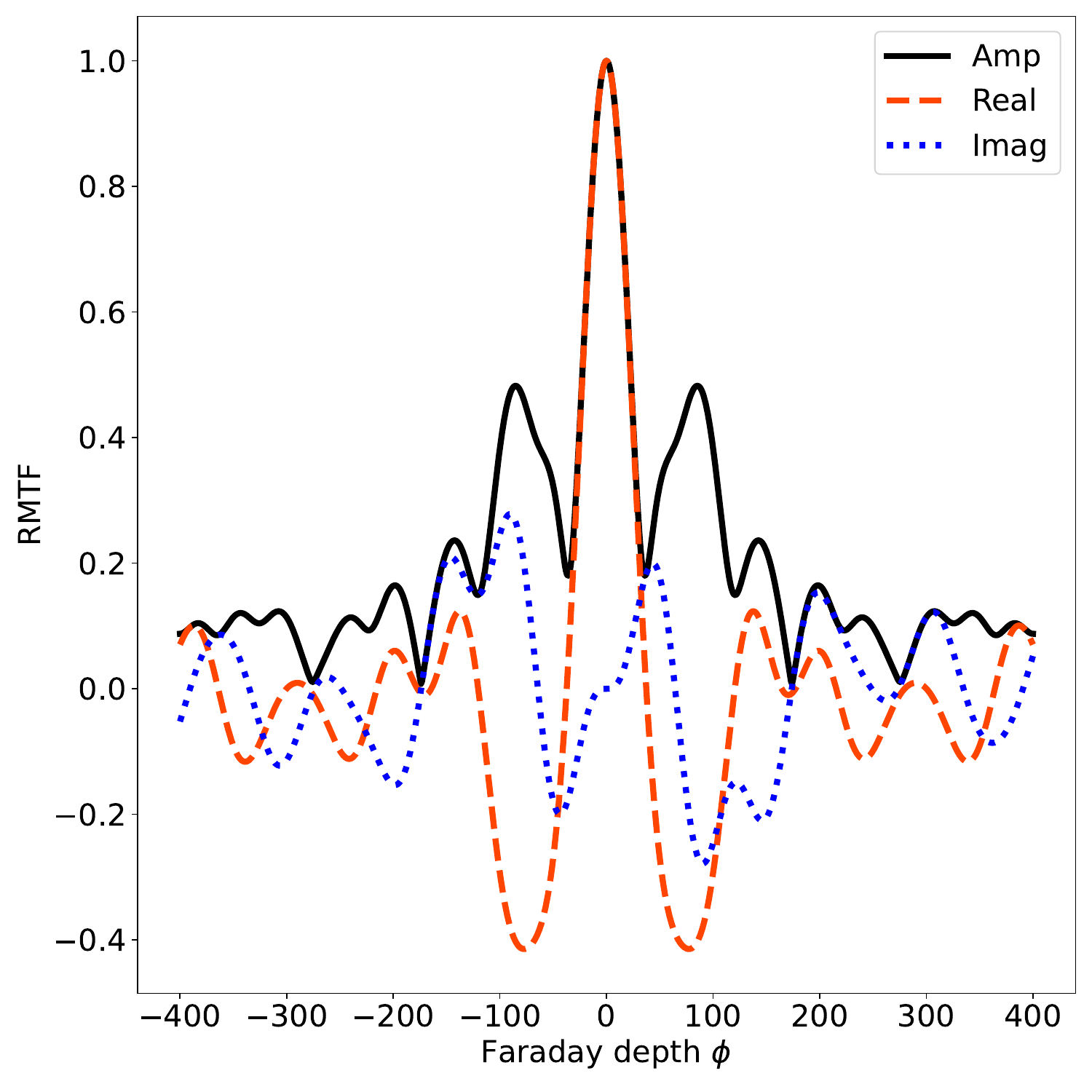}
    \caption{The RMTF of our data with a resolution (FWHM) of 41 rad m$^{-2}$. The solid, dashed, and dotted lines represent the amplitude, real and imaginary components, respectively. The high sidelobes are a result of missing frequencies that were flagged out due to RFI.}
    \label{fig:rmtf}
\end{figure}

On the other hand, the expected Faraday depth resolution, $\delta\phi$ (the full width half maximum, FWHM, of the RMTF), is dependent on the bandwidth in $\lambda^2$ space,  $\Delta(\lambda^{2}) = \lambda_{\text{max}}^{2} - \lambda_{\text{min}}^2$, as follows \citep{Schnitzeler2009}:
\begin{equation}
    \label{eqn:res_max}
    \delta\phi = \frac{3.8}{\Delta(\lambda^{2})}.
\end{equation}
The RMTF resolution for our data is $\sim$ 41 rad m$^{-2}$, and its structure is illustrated in Fig.~\ref{fig:rmtf}. Due to a large gap in the data (resulting from flagging frequencies that were significantly affected by RFI), our RMTF exhibits large sidelobes.
We performed RM-CLEAN \citep{heald2009a} on the derived Faraday spectra to extract more accurate values of RM. The RM-CLEAN algorithm is similar to the standard imaging CLEAN algorithm \citep{hogbom1974}. Peaks in the Faraday spectrum are iteratively identified, scaled, and subtracted until a minimum threshold or a maximum number of iterations is achieved. The clean components are then convolved by an idealised RMTF, e.g. a Gaussian function of width equivalent to that of the RMTF, resulting in a `cleaned'  Faraday spectrum. Besides deconvolution, RM-CLEAN attempts to mitigate problems arising from a large first sidelobe that could mimic a peak and cause confusion in cases where polarized emission is detected at multiple depths along a line-of-sight \citep{Heald2009}.

\section{Results}
\subsection{Total Intensity}
\label{sec:total-intensity-features}

\begin{figure*}
    \centering
    \includegraphics[width=2\columnwidth]{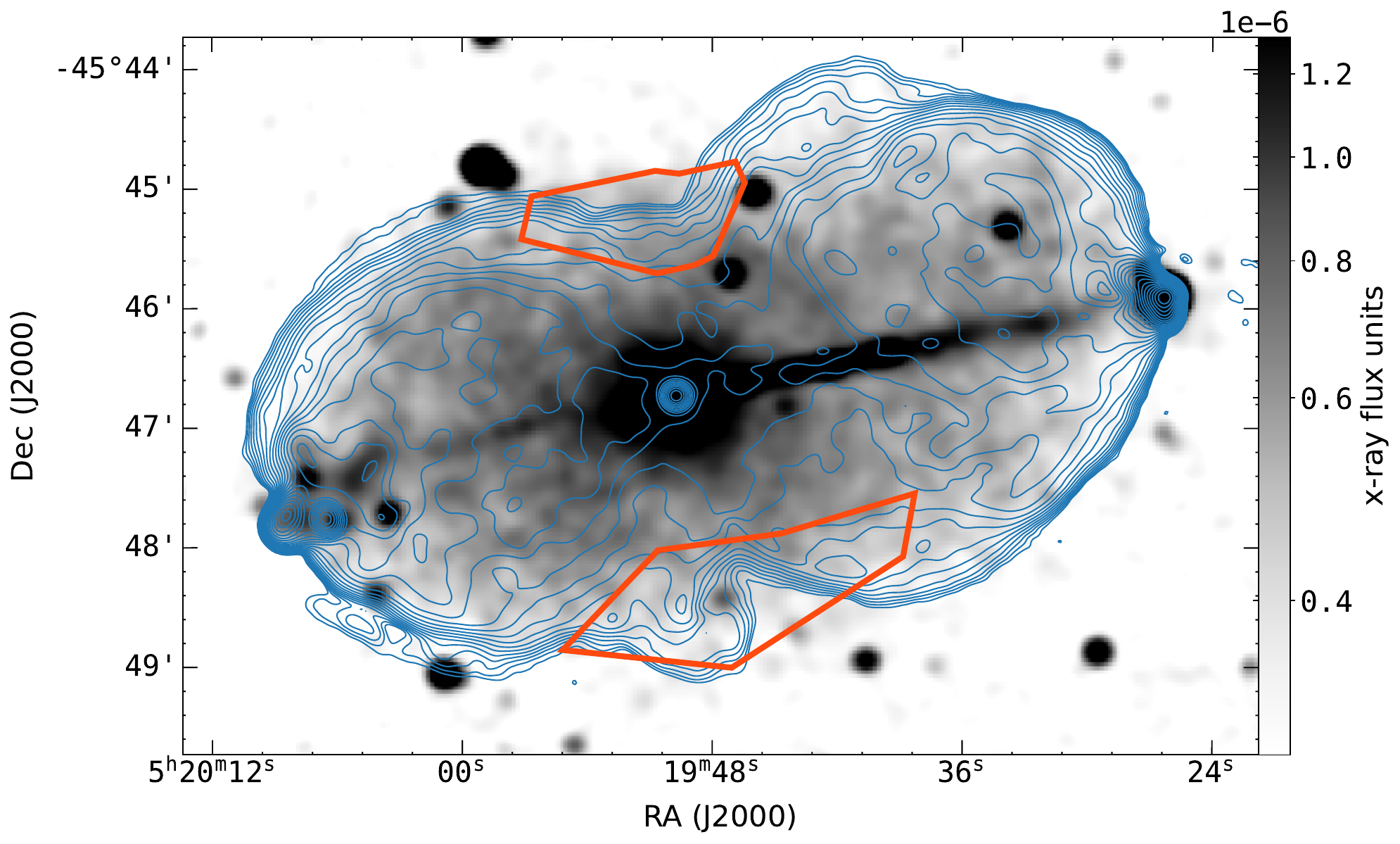}
    \caption{The Pictor A Chandra X-ray image from \citet{Hardcastle2016} made with 0.5--5.0 keV passband data (in greyscale) overlaid with radio emission contours from our data (in blue) at $\sim$7.5\arcsec resolution. Contours start from 0.6 mJy beam$^{-1}$ similar to Fig 1 of \citet{Hardcastle2016}, and separated by a factor of $\sqrt{2}$. The X-ray image was convolved to a resolution similar to our radio data and is shown in a log scale to emphasize the low-level emission. The orange polygons show a region previously identified by \citet{Hardcastle2016} as a region with missing radio flux emission. This radio emission is visible in the image. The X-ray jet is also excellently aligned with the radio jet.}
    \label{fig:missing-sync}
\end{figure*}

\subsubsection{Hotspots}
\label{sec:hotspots}
In Fig.~\ref{fig:pica}, we show the total intensity map of Pictor A obtained from our L-band data at 7.5\arcsec resolution -- the highest possible resolution for our observations. Our data show a single hotspot at the extremity of the western lobe, herein the western hotspot, and two hotspots at the extremity of the eastern lobe, herein the eastern hotspot (EHS). The presence of these hotspots is consistent with previous observations of Pictor A. 
The WHS is exceptionally bright, with a peak brightness of 7.2 Jy beam$^{-1}$ ($\sim$118 mJy arcsec$^{-2}$), while the inner EHS shows a peak brightness of 1.62 Jy beam$^{-1}$ ($\sim$26.5 mJy arcsec$^{-2}$), and the outer shows 1.01 Jy beam$^{-1}$ ($\sim$ 16.5 mJy arcsec$^{-2}$). At our resolution, these hotspots and radio core remain unresolved.
    
According to the dentist drill model \citep{scheuer1982},  the inner hotspot is the primary hotspot (most recent), and the outer hotspot is the secondary hotspot (older). This model implies that the eastern jet was first directed towards the secondary hotspot and then later changed direction towards the primary hotspot. The change in the jet direction occurred on time-scales shorter than the time taken for the secondary hotspot to dissipate due to expansion losses. Perhaps this may explain the differences in peak brightness of the hotspots -- with the primary hotspot being brighter than the secondary -- but this inference remains inconclusive and subjective.

From the spectral index analysis presented in section~\ref{sec:spi}, the spectral index of the primary hotspot (inner) is $-0.76 $ and of the secondary (outer) hotspot is $-0.79$ with standard deviations of 0.01 and 0.021 respectively (we define the relationship between spectral index, $\alpha$, and integrated flux density, $S$, as $S_{\nu} \propto \nu^{-\alpha}$). Assuming that the spectral index provides a good representation of the emission's age, the spectral index of the hotspots suggests that the primary hotspot is younger than the secondary -- consistent with the prediction by the Dentist drill model. However, the difference in the spectral indices is minimal -- comparable to the spread (standard deviation) in the spectral index distribution within the hotspots; thus, it cannot provide strong evidence for the age difference between them.
     
\subsubsection{Lobes}
\label{sec:lobes}
The notably round radio lobes are also visible from Fig.~\ref{fig:pica} are consistent with those seen in previous studies, except that our data reveals more diffuse emission due to higher sensitivity. The eastern lobe has an integrated flux density of 20.9 Jy, and the western lobe of 20.5 Jy, leading to a flux density lobe ratio of the eastern to western lobe of $\sim 1.02$. Radio filaments within the lobes are also observed similar to those seen by \citet{perley1997}. These filamentary structures are known to form as a result of shearing and strengthening of weak magnetic fields due to the dynamo amplification effect in filamentary regions \citep[see][for MHD simulation and details of the dynamo effect]{clarke96,donnert2018}. 
The `un-jetted' lobe has also been shown to be associated with a slightly steeper spectral index (see also our spectral indices in Sec~\ref{sec:spi}) and is thought to be spatially smaller owing to higher pressure from its denser environment \citep{perley1997}. This high pressure was speculated to enhance its magnetic field, leading to a higher observed spectral age \citep{liu1991, liu_correlated_1991}. Similar to \citet{perley1997}, we find the length ratio of the western to the eastern lobe to be 1.19 and a comparable ratio between the widths, indicating that the western lobe is larger. However, this could also be a projection effect.

In Figs.~\ref{fig:pica} and~\ref{fig:missing-sync}, we also note a region akin to what \citet{Carilli1996} call `radio plumes'. In Pictor A's case, plumes appear midway between the lobes on both the far south and north of the lobe emission. These plumes appear as emission `spilling' from the central lobe emission. \citet{Carilli1996} attribute plumes to the displacement of radio-emitting plasma away from the galactic core due to pressure gradients from the intracluster medium.
\citet{Hardcastle2016} also pointed out that diffuse X-ray emission appeared to extend beyond the visible lowest VLA data radio contour at 0.6 mJy \citep[see fig. 1 of][]{Hardcastle2016} at the `waist' of the source. This contradicts expectations since the X-ray lobe emission is suggested to be inverse Compton (IC) in origin; thus, we expect to see the spatial superposition of the X-ray and radio. The missing radio emission was attributed to either 1) the limited sensitivity of the VLA at the time, 2) the X-ray emission being thermal in origin within these regions, or 3) the radio emission being too faint or of steep-spectrum.

Fig.~\ref{fig:missing-sync} shows an X-ray image of Pictor A in greyscale from \citet{Hardcastle2016} overlaid with contours from our radio data. The lowest radio contour here is 0.6 mJy beam$^{-1}$, with consecutive contours separated by a factor of $\sqrt{2}$. The previously unseen radio emission is visible (marked by the orange polygons), consistent with the suggestion that X-ray lobe emission is predominantly of IC/CMB origin. However, \cite{Hardcastle2016} pointed out that the soft-thermal emission cannot be ruled out.
     
\subsubsection{Radio Jets}
\label{sec:jet}
Pictor A's jets are remarkably visible in the X-rays \citep[e.g. see Fig.~\ref{fig:missing-sync} and][]{Wilson2001,Hardcastle2016}, the western jet clearly shows to extend from the core and terminates at the hotspot ($\sim$ 174 kpc). The jets emission was determined to be of synchrotron origin \citep{Hardcastle2016}. While the western radio jet was slightly noticeable in the study of \citet{perley1997} using the VLA data, the authors remarked that it was very faint and was best visible on a TV monitor. We show for the first time in Fig.~\ref{fig:pica} a more pronounced western radio jet from the radio nucleus to the WHS. Although this radio jet remains unresolved, it is collimated and consists of a series of elongated knots, similar to those observed by \citet{Carilli1996} on the jets of Cygnus A. These knots were also partially observed in optical frequencies using the Hubble Space Telescope (HST) by \citet{gentry2015}. This radio jet has a position angle (as per the IAU convention) of $\sim281\pm1^{\circ}$ and runs from the centre of the radio core to about 1.93\arcmin, after which its emission diffuses into the western lobe. Though the jet becomes indistinguishable at this point, tracing it along this position angle leads directly to the WHS.

By contrast, the counter-jet is still not detected in radio images. Although discernible in the X-ray, it seems to diffuse into the lobe emission near the nucleus and hotspots \citep{Hardcastle2016}. Our findings are consistent with those of the previous studies; there is no discernible counter-jet. 
This apparent jet single-sidedness in a radio galaxy like Pictor A's (in Fig.~\ref{fig:pica}) could result from the Doppler boosting effect that makes the jet closer to the observer appear brighter and more visible.
\begin{figure}
    \centering
    \includegraphics[width=\columnwidth]{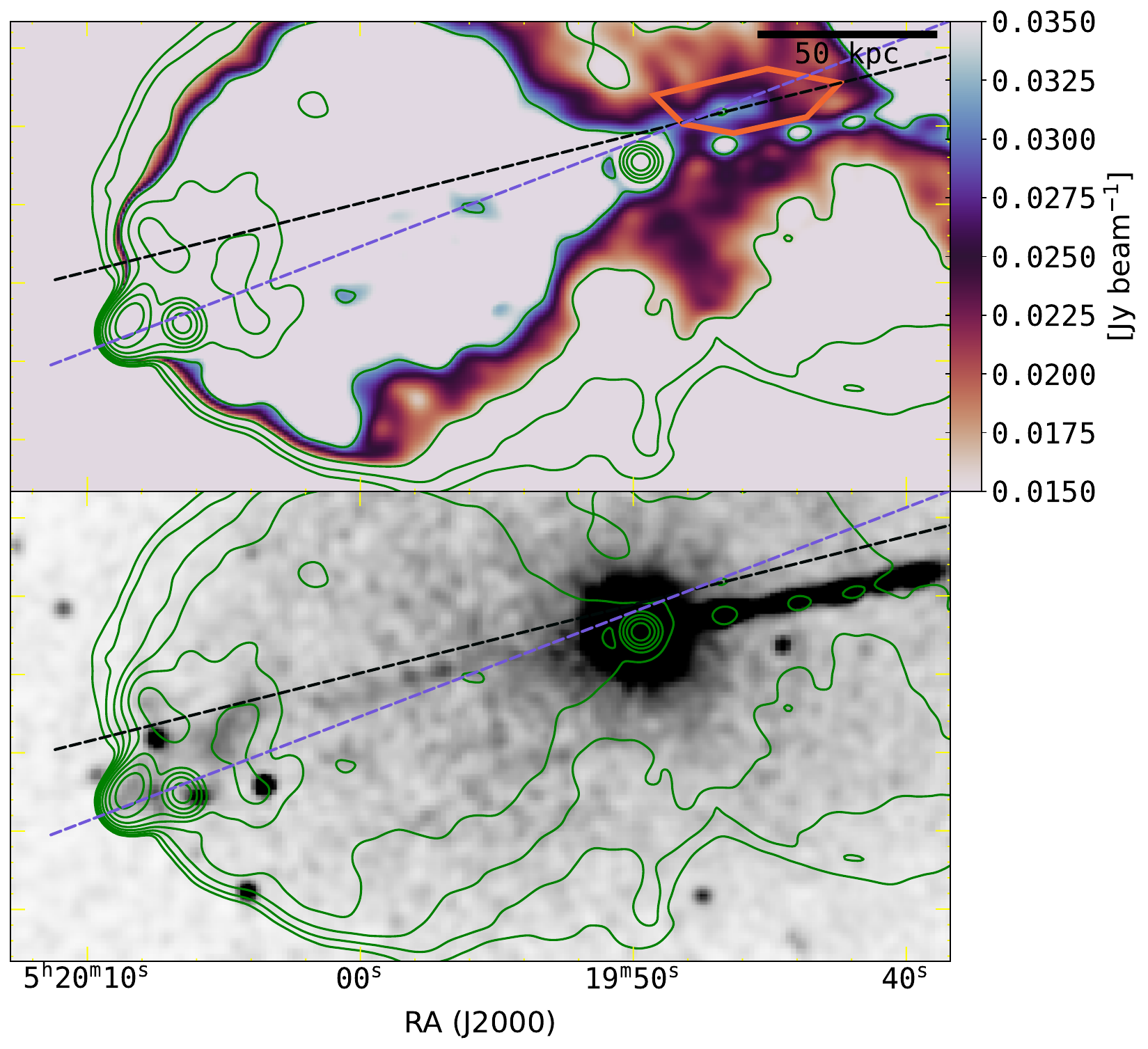} 
    \caption{A novel collimated feature reminiscent of a radio jet observed from our data. The top panel highlights this feature marked by the orange polygon. The green contours show the radio emission, while the straight lines show the possible extension of the paths of this feature. The bottom panel shows the X-ray image superposed with the same radio contours as the top panel. The straight lines are also drawn at the same position.}
    \label{fig:jetlike}
\end{figure}

Fig.~\ref{fig:jetlike} shows a feature near and North-West of the AGN reminiscent of a radio jet due to its semblance in total intensity and collimation to the jet's emission rather than the neighbouring lobe emission. Since this feature is not directly aligned with the WHS and the core, it is difficult to tell conclusively. Nevertheless, we drew two possible lines based on the knot extension to trace its extension towards the lobe's extremity. To our knowledge, this feature was not seen in previous radio images and may be of interest in the future.

\subsubsection{Spectral Index}
\label{sec:spi}
As a result of the excision of data affected by RFI, we generated our spectral index map using a frequency range smaller than that of the MeerKAT telescope's L-band, 0.89--1.64 GHz, using the \texttt{spimple} tool. We consider $0 > \alpha \geq -0.7$ as flat and $\alpha < -0.7$ as steep. 
\begin{figure}
    \centering
    \includegraphics[width=\columnwidth]{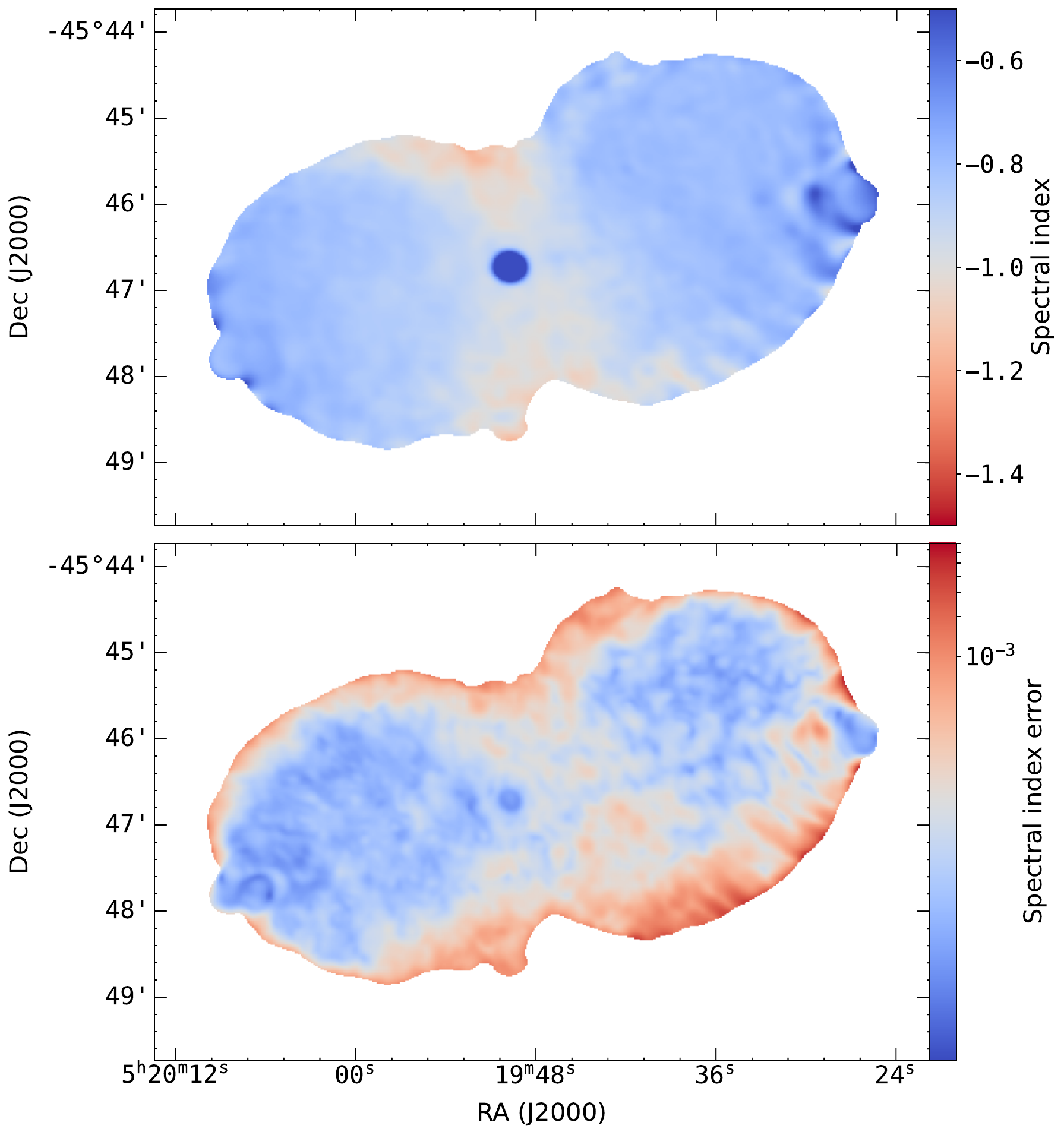}
    \caption{The spectral index map generated using channelised images between 0.88 and 1.65 GHz at a resolution of 11\arcsec by 10\arcsec (top panel). The spectral index colour bar ranges from $-0.5$ to $-1.2$. Spectral steepening is seen towards the waist of Pictor A, while the core shows a flat spectral index. The bottom panel shows errors associated with spectral index fitting, whereby higher errors are associated with the lobe edges.}
    \label{fig:spi-contours}
\end{figure}
The top panel in Fig.~\ref{fig:spi-contours} reveals that both lobes have fairly similar spectral indices, with a mean of $\left\langle\alpha\right\rangle = -0.86$. Furthermore, the EHS pair displays a steeper spectral index of $\left\langle\alpha\right\rangle = -0.76$ and $-0.79$ for the inner and outer hotspots, respectively, compared to that of the WHS $\left\langle\alpha\right\rangle = -0.64$, while the radio core has $\left\langle\alpha\right\rangle =-0.48$. These values are summarised in Table ~\ref{tab:spi-regions}. In that table, we also indicate the number of independent beams that fit within the specified region of Pictor A rounded off to the nearest integer. They were obtained by dividing the region area by the beam's area. 
\begin{table}
    \centering
    \caption{Mean spectral indices for selected regions within Pictor A. The mean frequency is 1.21 GHz. The last column of this table shows the number of independent beams that fit in the specified regions within the source.}
    \label{tab:spi-regions}
    \begin{tabular}{lccc} 
        \hline
        Region & Mean $\alpha$ & Std Deviation $\alpha$ & Beams\\
        \hline 
            E-lobe    &   $-0.88\pm 2.0 \times 10^{-4}$   &    $0.09$ & 291 \\
            W-lobe    &   $-0.85\pm 3.7 \times 10^{-4}$   &    $0.08$ & 382 \\
            E-Hotspot (inner)  &   $-0.76\pm 6.4 \times 10^{-6}$   &    $0.010$ & 2 \\
            E-Hotspot (outer)  &   $-0.79\pm 3.0 \times 10^{-5}  $   &    $0.021$ & 3 \\
            W-Hotspot      &   $-0.64\pm 2.0 \times 10^{-4}$   &    $0.085$ & 11 \\
            Core     &   $-0.48\pm 2.0 \times 10^{-5}  $   &    $0.27$ & 5 \\
            Plume   &   $-1.02\pm 7.0 \times 10^{-4}$   &    $0.074$ & 38 \\
            Waist    &   $-0.94\pm 3.1 \times 10^{-4}$   &    $0.12$ & 223 \\
        \hline
    \end{tabular}
\end{table}
In contrast, the source's waist region displays significant spectral steepening away from the core and towards its periphery along the north and south edges. These spectral indices align well with the observations of \citet{perley1997}, who stated a value of $\left\langle\alpha\right\rangle = -0.8$ for the lobes and noted that the eastern lobe has a steeper spectrum than the western lobe. However, we found a maximum of $\alpha \sim -1.3$ along the waist compared to \citeauthor{perley1997}'s $-1.7$. Errors associated with our spectral index fitting are shown in the bottom panel of Fig.~\ref{fig:spi-contours}. 
These are the random errors propagated from deconvolution and are calculated as:
\begin{equation*}
    \alpha_{\text{err}} = \sqrt{\left(\frac{\Delta I_{0}}{I_{0}}\right)^{2} + \left(\frac{\Delta I_{1}}{I_{1}}\right)^{2}},
\end{equation*}
where $\Delta I_{0}$ and $\Delta I_{1}$ are the RMS errors measured from the first and second Taylor-coefficient maps.
Higher error areas appear to be associated with the lobe's edges. We attribute this to low signal-to-noise along those regions coupled with edge effects. The `stripy' structure on both maps results from residual calibration artefacts.

This type of spectral variation is typical of FR-II sources. Usually, the hotspots and core exhibit a flatter spectrum, while the lobe exhibits a generally steeper spectrum. For Pictor A, this is true except for the eastern hotspots, with spectral indices comparable to the lobes. Flat spectral indices indicate the presence of younger electrons, and conversely, a steeper spectral index indicates older electrons. Our map shows that the age of electrons increases away from the core in the N-S direction.

Spectral ageing is thought to occur due to more energetic electrons radiatively losing their energy faster in the absence of acceleration, resulting in spectral cooling and a steepened spectrum. Models such as the Kardashev--Pacholczyk (KP) and Jaffe--Perola (JP) have been proposed to predict the energy spectral shapes resulting from spectral ageing \citep[see][for a summary and references therein for a complete treatment]{harwood2013}. These models assume a fixed magnetic field strength across the source and a single injection electron distribution at the jet termination points. However, to factor in a more realistic magnetic field structure, \citet{tribble_radio_1993} introduced more advanced spectral ageing models that assume a Gaussian randomly varying magnetic field that allowed electrons to diffuse with the varying field strengths. Identifying the more accurate model without deeply understanding the underlying physical processes is difficult. Unfortunately, the relatively limited bandwidth of our current data confines the extent to which we can probe the spectral behaviour of this source. Thus, studies with a broader bandwidth coverage will be invaluable in better characterising the spectral indices of Pictor A.

\subsection{Polarimetry}
\label{sec:polarimetry-results}
\subsubsection{Linear Polarized Intensity and Fractional Polarization}
\begin{figure}
    \centering
    \begin{minipage}[b]{1\linewidth}
        \centering
        \includegraphics[width=\columnwidth]{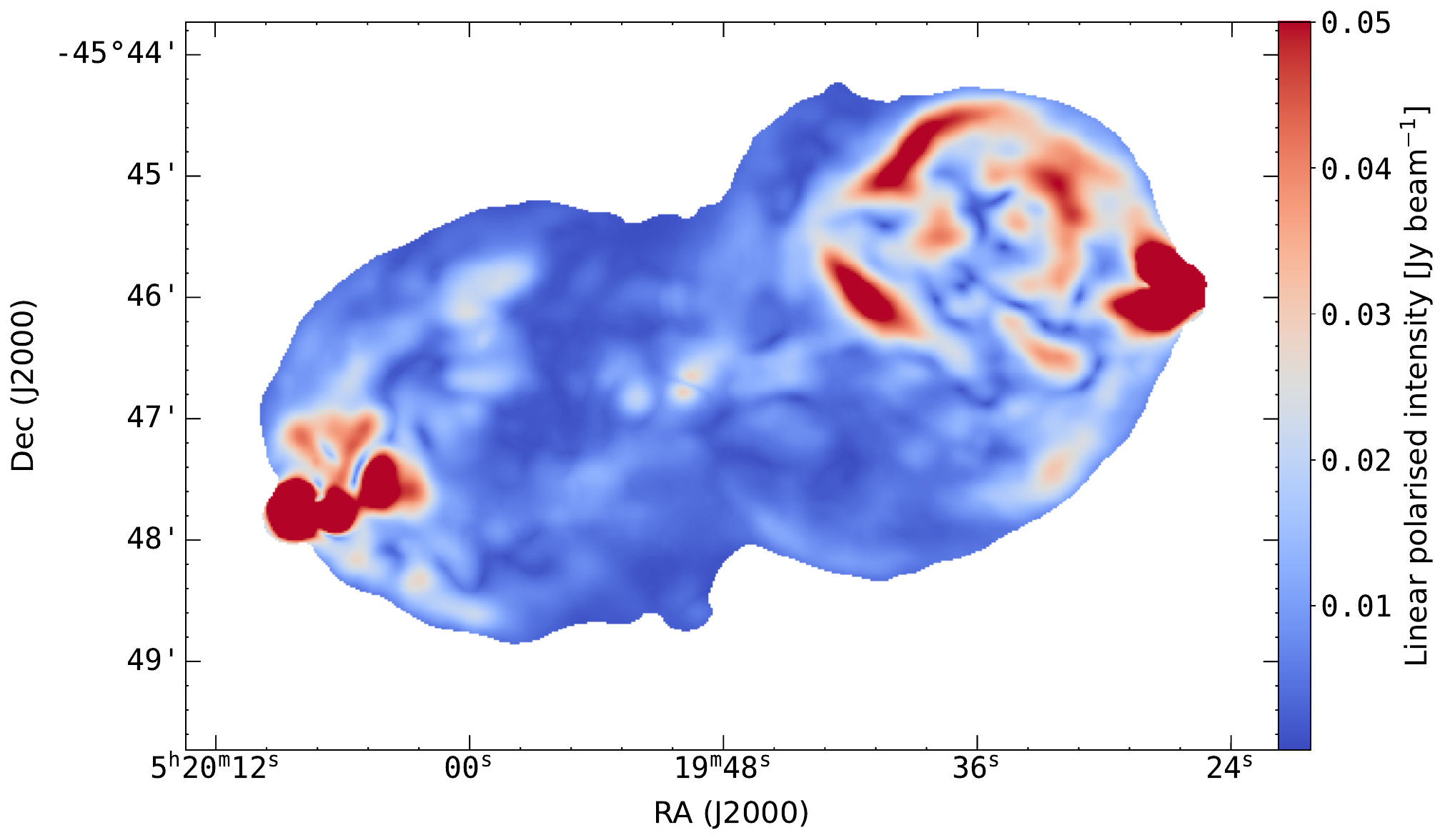}
    \end{minipage}
    \begin{minipage}[b]{1\linewidth}
        \centering
        \includegraphics[width=\columnwidth]{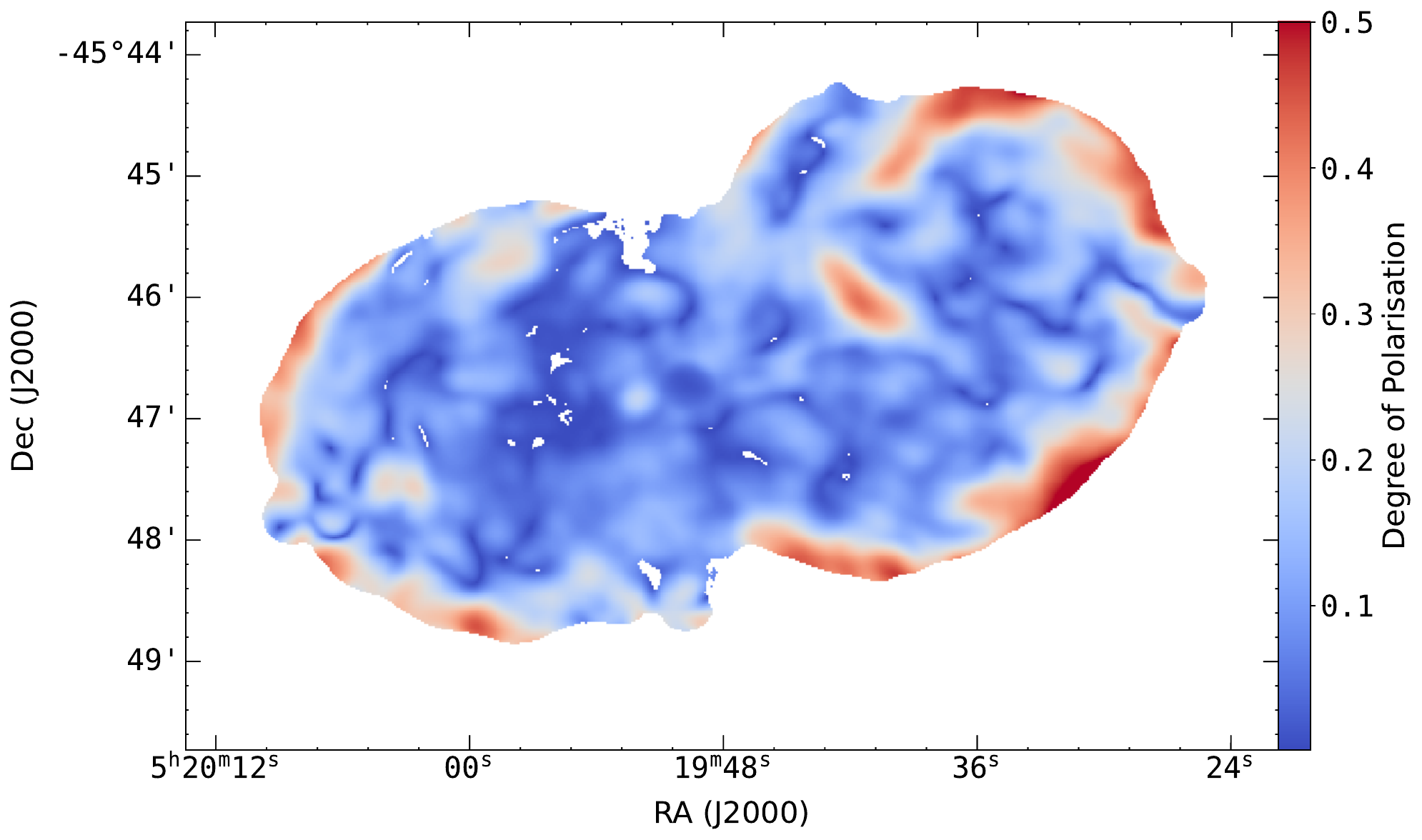}
    \end{minipage}
  \caption{The linear polarized intensity across Pictor A (top panel) and the degree of polarization (bottom panel) at 1.65 GHz, whereby the hotspots and radio filaments show the highest amounts of linear polarization. Both images are at 11\arcsec by 10\arcsec in resolution, and the top panel's units are in Jy beam$^{-1}$.}
  \label{fig:linear-poln}
\end{figure}

Fig.~\ref{fig:linear-poln} shows the linear polarization (top) and degree of polarization across the lobes of Pictor A at 1.65 GHz and 11\arcsec by 10\arcsec resolution. We compute the polarized brightness using Eqn.~\ref{eqn:pol-amp}. High linear polarization in Pictor A is found at the hotspots, within the western lobe's filamentary structure and close to the EHS. This is similar to the findings of \citet{perley1997}. There seems to be no correlation between filaments in the total intensity image and those in the polarized brightness distribution. 

The degree of polarization (fractional polarization), $p$, was given as $|P|/\mathit{I}$. We obtained images of $p$ from the sub-band $\mathit{I}$ maps as described in Sec.~\ref{sec:polarimetry}. No Ricean bias correction was made for any of the images.
High degrees of polarization are seen around the edges of Pictor A, and a single filamentary structure located at around RA $5^{\mathrm{h}} 19^{\mathrm{m}} 41^{\mathrm{s}}$ and Dec $-45^{\circ} 46^{\prime} 00\arcsec$. In this figure, the colour bar is representative of the degree of polarization. Levels of the fractional polarization over most of the lobes' regions are below 10 per cent, whereby the western lobe demonstrates more regions with higher fractional polarization values than the eastern lobes.
As noted by \citeauthor{perley1997}, there seems to be no correlation between the regions of higher total intensity and fractional polarization. Their observations at 7.5\arcsec resolution showed that some regions displayed up to 70 per cent fractional linear polarization. From our data in Fig.~\ref{fig:linear-poln}, a few regions within the western lobe also appear to show a degree of polarization approaching 60 per cent. 

\subsubsection{Line-of-Sight Fractional Polarization}
\begin{figure*}
    \centering
    \includegraphics[width=2\columnwidth]{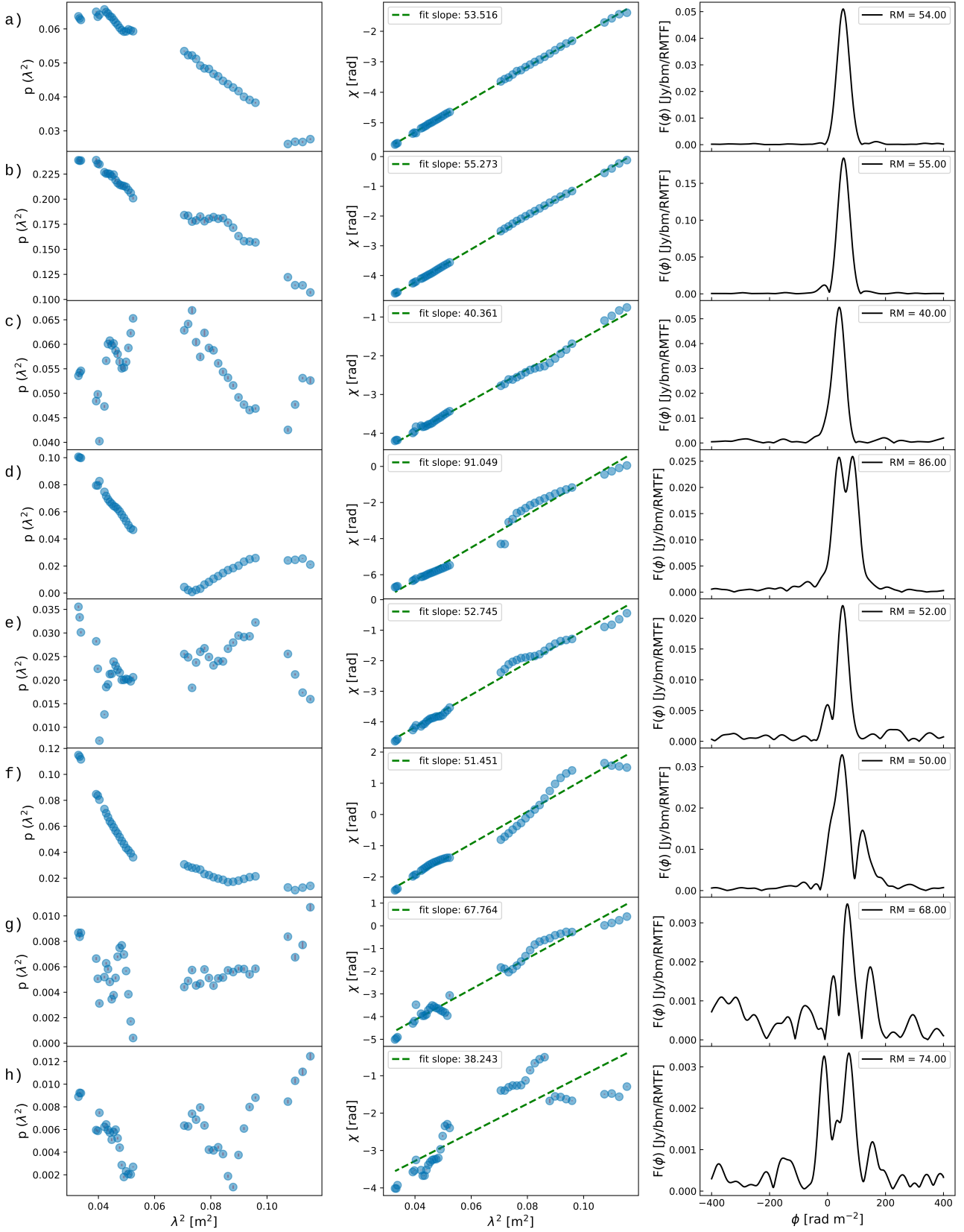}
    \caption{The polarization behaviour of example lines-of-sight across the lobes of Pictor A. Left column: fractional polarization vs $\lambda^2$. Middle column: polarization angle vs $\lambda^2$ superimposed by a linear fit (green dashed line). Right column: the RM-cleaned Faraday spectra are represented. RM values shown on the legend of this column indicate the position in Faraday depth of the Faraday spectrum's highest peak. The RMTF is shown in Fig.~\ref{fig:rmtf}.}
    \label{fig:los-pol}
\end{figure*}

Fig.~\ref{fig:los-pol} shows polarization structures for the individual lines-of-sight (LoS) across the lobe of Pictor A. We have 2389 lines-of-sight across Pictor A, spatially separated by approximately 6\arcsec. These lines-of-sight were chosen using the criteria of having a signal-to-noise ratio (SNR) above 50 in total intensity image at 1 GHz. For illustration, we can only show eight of the lines-of-sight. The left column is the degree of polarization as a function of wavelength-squared, and the middle column consists of the unwrapped polarization angle as a function of wavelength-squared in blue points and the fitted linear model in green dashed lines. The third column is the cleaned Faraday spectrum (cleaned FDF) in black superimposed by the dirty FDF in red dashed line. In this plot, the top three rows highlight examples of lines-of-sight with a single peak in the FDF, the fourth and fifth rows show examples with double peaks, and the last three rows show examples with multiple peaks. The lobes of Pictor A are dominated by single-peaked FDF. 

\begin{figure*}
    \centering
    \includegraphics[width=2\columnwidth]{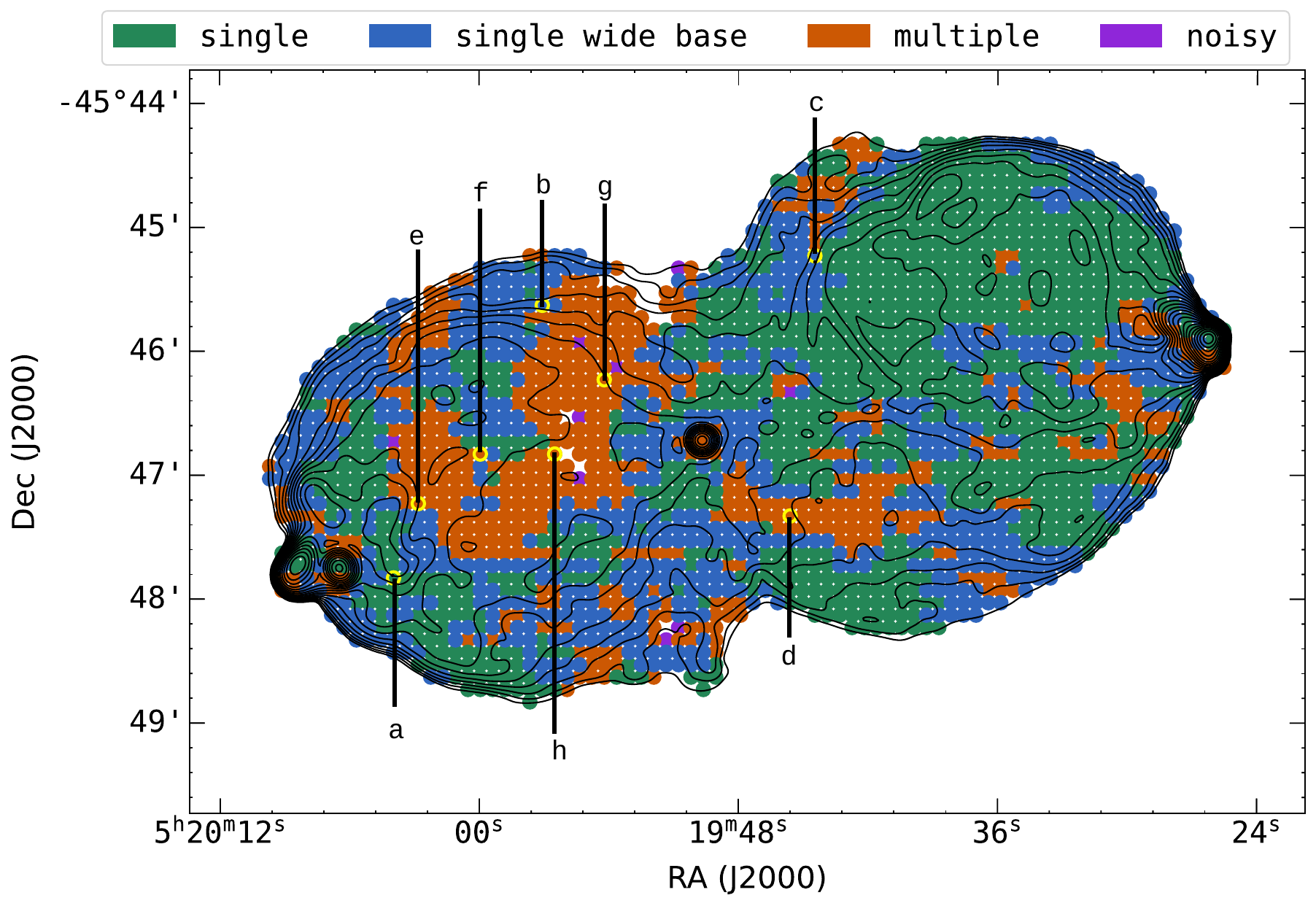}
    \caption{Locations of the various categories of LoS Faraday spectra exhibited by Pictor A. Most of the lines-of-sight show Faraday spectra with a single dominant peak (shown in green). In other cases, this single peak has a wide base or an additional smaller peak on either side (in blue). This source also exhibits multiple peaks in some regions (represented by orange). Noisy lines-of-sight are represented by purple. Lines-of-sight outlined in yellow are those exemplified by Fig.~\ref{fig:los-pol}, and are labelled accordingly.}
    \label{fig:spectra-plot-cat}
\end{figure*}

Quantitatively, $\sim$43.5 per cent of the LoS exhibited Faraday spectra with a single dominant peak such as those illustrated in row one of Fig.~\ref{fig:los-pol}. Additionally, we found that for these single-peaked LoS, $\sim$78 per cent showed a maximum fractional polarization greater than 10 per cent, while $\sim$33 per cent showed fractional polarization of more than 20 per cent. On the other hand, $\sim$32.6 per cent of the single-peaked LoS spectra showed a wide base and/or smaller peaks on either side (such as those of rows two and three of Fig.~\ref{fig:los-pol}). We also found that $\sim$23.5 per cent of the spectra showed double peaks (e.g. those in rows 4, 5 and 6 of Fig.~\ref{fig:los-pol}), while spectra of some a few of the boundary regions associated with more than two peaks (rows 7-8 in the same figure). The remaining $\sim$0.3 per cent of Los spectra were deemed to be noisy. Shown in Fig.~\ref{fig:spectra-plot-cat} are the locations of these various categories of spectra, whereby LoS spectra with single, single wide base, and multiple peaks are represented by green, blue and orange respectively. Noisy regions are represented by purple. Fig.~\ref{fig:spectra-plot-cat} also highlights the positions of the lines-of-sight exemplified in Fig.~\ref{fig:los-pol} using yellow outlines. These LoS have also been labelled accordingly. Pictor A's lines-of-sight are available at \url{https://pica.ratt.center/} displayed using the \texttt{PolarVis} tool \citep{andati2023}, which allows the visualization of various diagnostic plots associated with the specific lines-of-sight.\footnote{\url{https://github.com/Mulan-94/polarvis}} The lines-of-sight and their corresponding interactive plots were generated using \texttt{Scrappy}, another \texttt{Python}-based tool that semi-automates generation of the LoS depending on a user's specifications.\footnote{\url{https://github.com/Mulan-94/scrappy}} We probed further into regions showing RM gradients and double peaks to determine if these effects were real. This is discussed in Sec.~\ref{sec:rm-gradients}.

The behaviour of fractional polarization vs $\lambda^2$ is quite complex, even for single-peaked spectra. In some cases, repolarization appears to occur at longer wavelengths with highly oscillatory structure. The position angle vs $\lambda^2$ mostly exhibits linearity except for the double and multi-peaked spectra. On the other hand, the double and multi-peaked Faraday spectra show a wide range of behaviour: in some lines-of-sight the components are well resolved, and in some cases, the components are blended. The spectra of the last row resemble that of a Burn-slab \citep{Burn1966} -- a region in which polarized emission and Faraday rotation occur simultaneously. It is evident that the polarization behaviour across the lobes of Pictor A is complex. 

\subsubsection{Depolarization}
\label{sec:depolarization}
Depolarization refers to decreased observed fractional polarization as a function of resolution or frequency. It can result from effects internal to the source (due to mixed thermal and magneto-ionized gas) or external to the source (such as beam depolarization) as highlighted by \citet{Burn1966, Sokoloff1998} and manifests more prominently at the longer wavelengths (lower frequencies). Depolarization can be quantified using a ratio of fractional polarization measurements at two different frequencies or resolutions -- i.e. the depolarization ratio. 

\begin{figure}
    \centering
    \includegraphics[width=1\columnwidth]{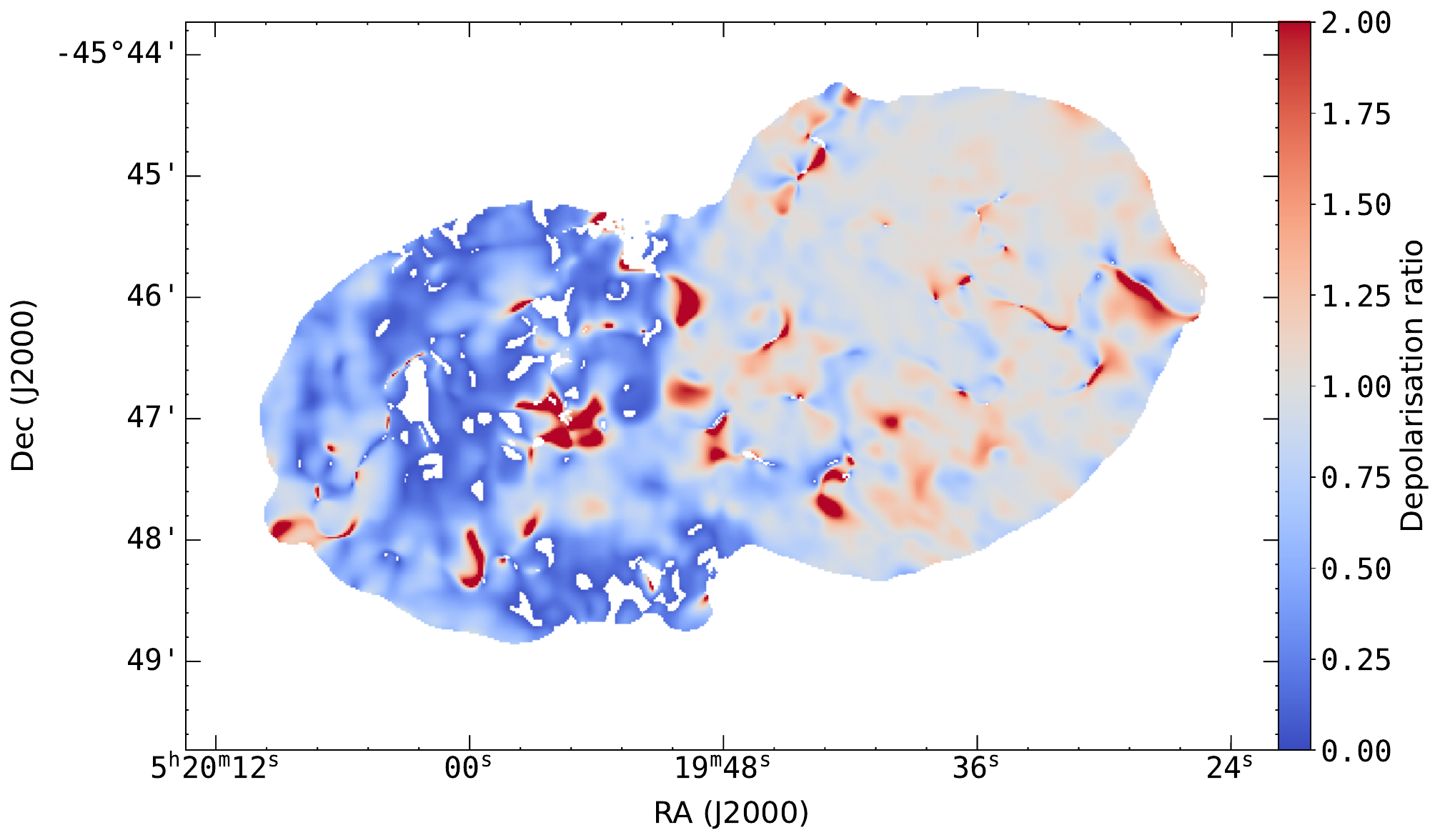}
    \caption{The depolarization ratio obtained by taking a ratio of the fractional polarization map at 0.88 GHz and 1.65 GHz, both at 11\arcsec by 10\arcsec resolution. The eastern lobe shows more depolarization than the western lobe. Here, values > 1 (towards bright red) show re-polarization, while those < 1 (towards deep blue) show depolarization. Only data where the depolarization error to depolarization ratio was > 0.6 are shown. Higher errors are associated with the eastern lobe.}
    \label{fig:depolarization}
\end{figure}
In Fig.~\ref{fig:depolarization}, we show the depolarization ratio map obtained by dividing the fractional polarization map at 0.88 GHz with a map at 1.65 GHz, both at 11\arcsec by 10\arcsec resolution. The depolarization is expected to be stronger at longer wavelengths; thus, a depolarization ratio of $<$ 1 (the bluer) suggests a depolarized region, while $>$1 (redder) suggests a re-polarizing region. Here, we show only regions where the fractional depolarization error\footnote{I.e. the standard propagated depolarization error ratio divided by the depolarization ratio.} was less than 0.6. We find that the eastern lobe shows more depolarization than the western lobe, particularly in the extreme lobe region -- consistent with the results of \citet{perley1997}, with some patches of re-polarization. This map bears a striking boundary separating the two lobes. The depolarization or re-polarization may be caused by unresolved Faraday structures (beam depolarization) or internal mixing of thermal and ionized gas within the lobes. Our data are insufficient to determine the actual cause for the depolarization and re-polarization within the lobes of Pictor A. However, these depolarization ratios must be interpreted cautiously, particularly because of our narrow bandwidth and significant variations in fractional polarization vs $\lambda^2$ (left column of Fig.~\ref{fig:los-pol}). The re-polarization may not necessarily be real or true but a result of limited bandwidth. 
\begin{figure}
    \centering
    \includegraphics[width=1\columnwidth]{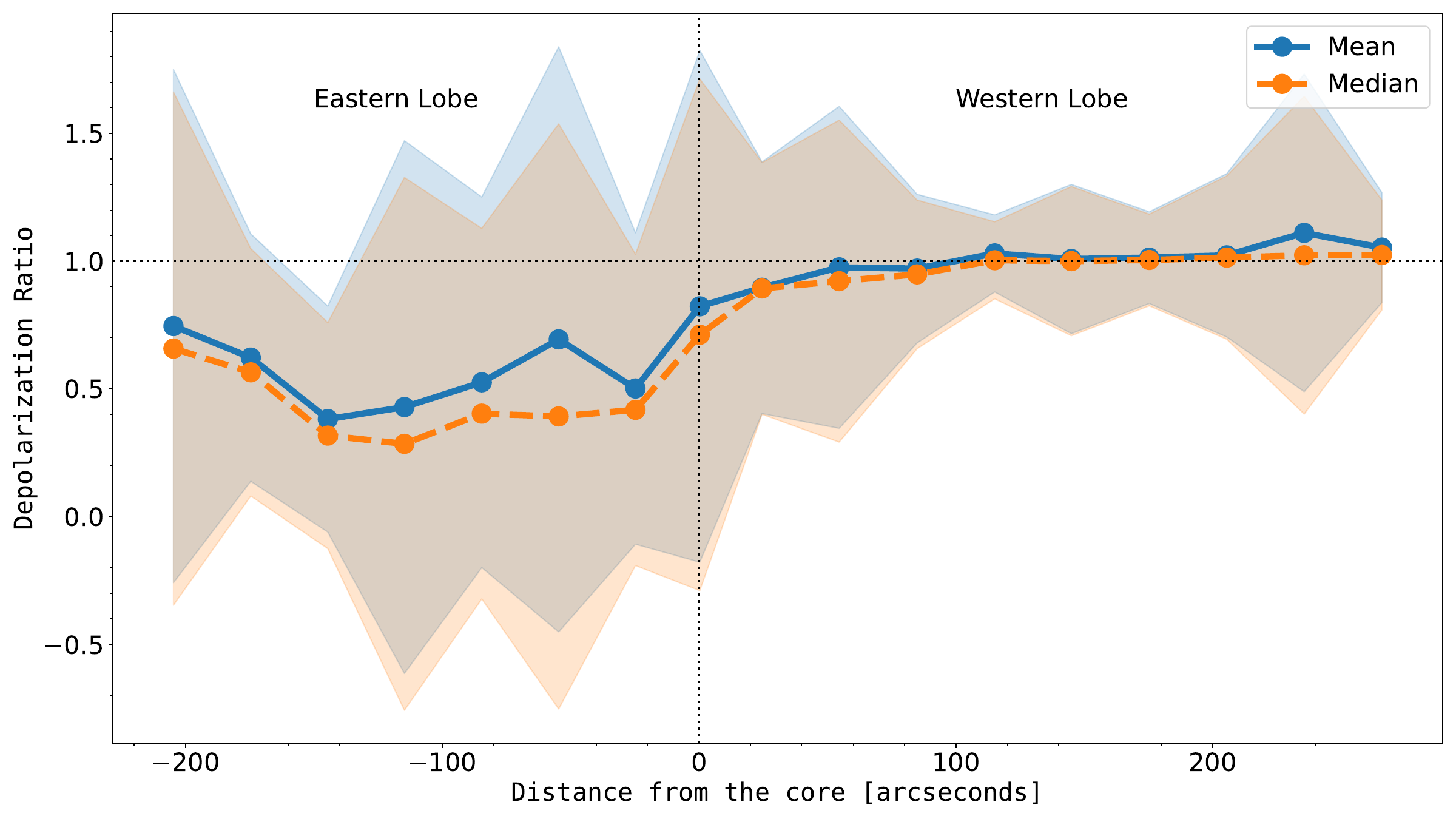}
    \caption{The binned depolarization ratio with distance from the radio core. Data points were obtained by subdividing Pictor A into rectangular regions of 30\arcsec in width along the jet axis and calculating the mean, median and standard deviation of data within those regions. The standard deviation is shown in the filled coloured patches, while the solid and dashed lines indicate the mean and median, respectively. For clarity, we also show a dotted line at a depolarization ratio of 1. Negative distances indicate a leftward direction from the radio core, while positive ones indicate a rightward direction. The vertical dashed line indicates the core's position.}
    \label{fig:depolarization-profile}
\end{figure}

Supporting depolarization profiles of both lobes are also shown in Fig.~\ref{fig:depolarization-profile} as a function of distance from the radio core, obtained by sub-dividing the source into box regions of 30\arcsec in width\footnote{Except for the bin encompassing the radio core, whereby the width is $\sim$ 20\arcsec. The radio core region was excluded from these bins.} along the jet axis and calculating a simple mean, median, and standard deviation from our depolarization map within the bins. The western lobe of this source shows consistent behaviour across the lobe, with values centred around a depolarization ratio of 1, indicating little to no depolarization. Contrastingly, the eastern lobe shows a more erratic spread with no particular trend with increasing core distance.

The asymmetric depolarization observed, where the jet-sided lobe depolarizes slower than its counterpart, has previously been defined as the Laing-Garrington effect \citep{Laing1988,Garrington1988}. This is an orientation effect that affects the lobe further from the observer because its emission undergoes more depolarization effects resulting from passing through a longer path length along the magneto-ionic halo of its host galaxy. Asymmetric depolarization is a common feature in FR-II sources \cite[e.g.][]{goodlet2004,ishwara-chandra1998}. 
    
Assuming the eastern lobe is further, its higher observed depolarization could result from external Faraday dispersion. This means that its emission may go through various turbulent magneto-ionized cells along a line-of-sight towards the observer, which causes depolarization and a resulting low polarized signal. Another possible cause of depolarization is the intermixing of thermal and synchrotron-emitting gas within the lobes. The significant depolarization of this lobe between 20 and 6 cm noted by \citeauthor{perley1997} means that low-frequency depolarization possibly further contributes to the observed lower average degrees of polarization in the eastern lobe visible in Fig.~\ref{fig:linear-poln}. As such, higher frequency observations are required to properly characterise the depolarization behaviour of Pictor A's lobes.

\subsubsection{Faraday Rotation Measures}
\label{sec:rm-distribution} 
\begin{figure*}
    \begin{minipage}[b]{1\linewidth}
        \centering
        \includegraphics[width=\columnwidth]{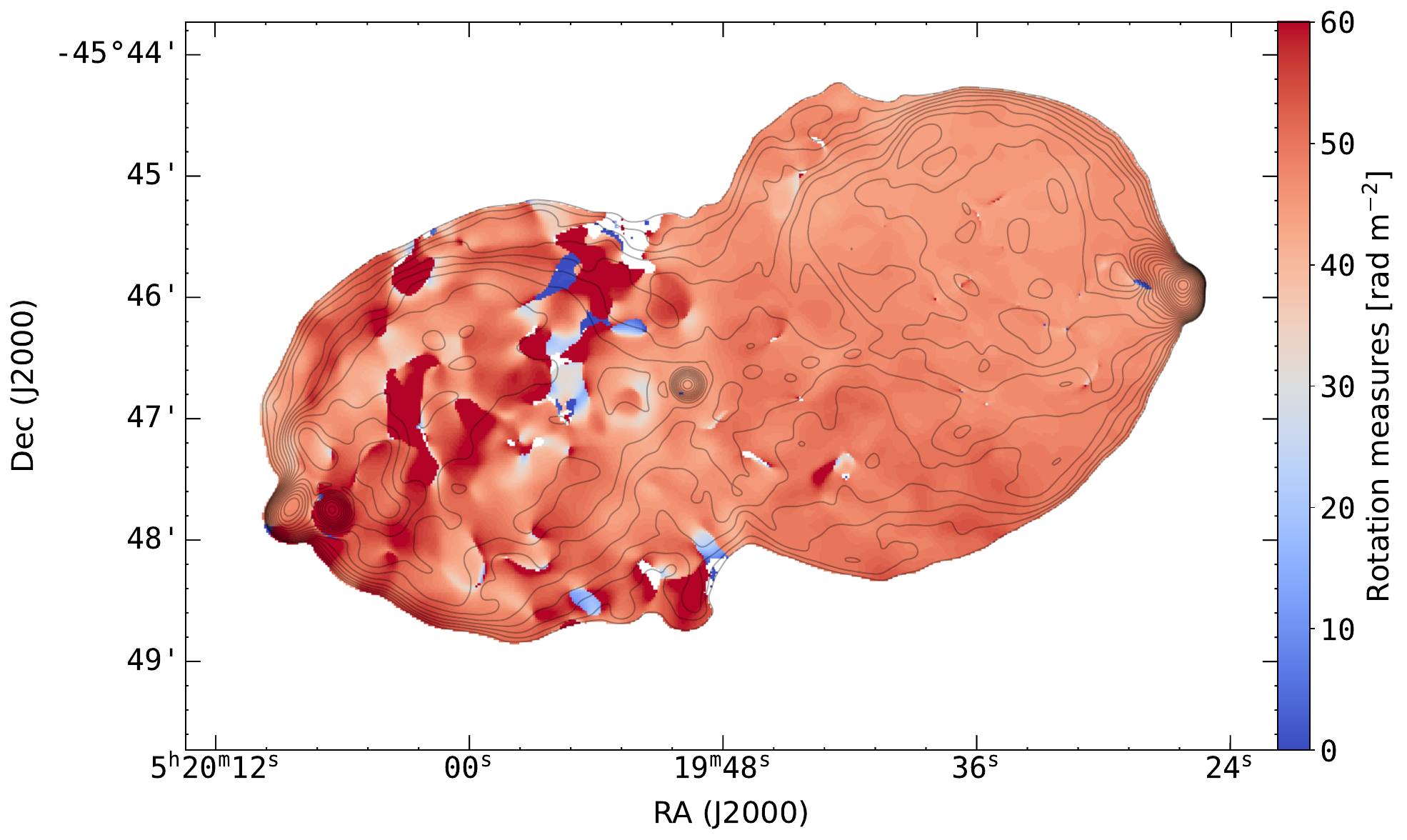}
    \end{minipage}
    \begin{minipage}[b]{1\linewidth}
        \centering
        \includegraphics[width=\columnwidth]{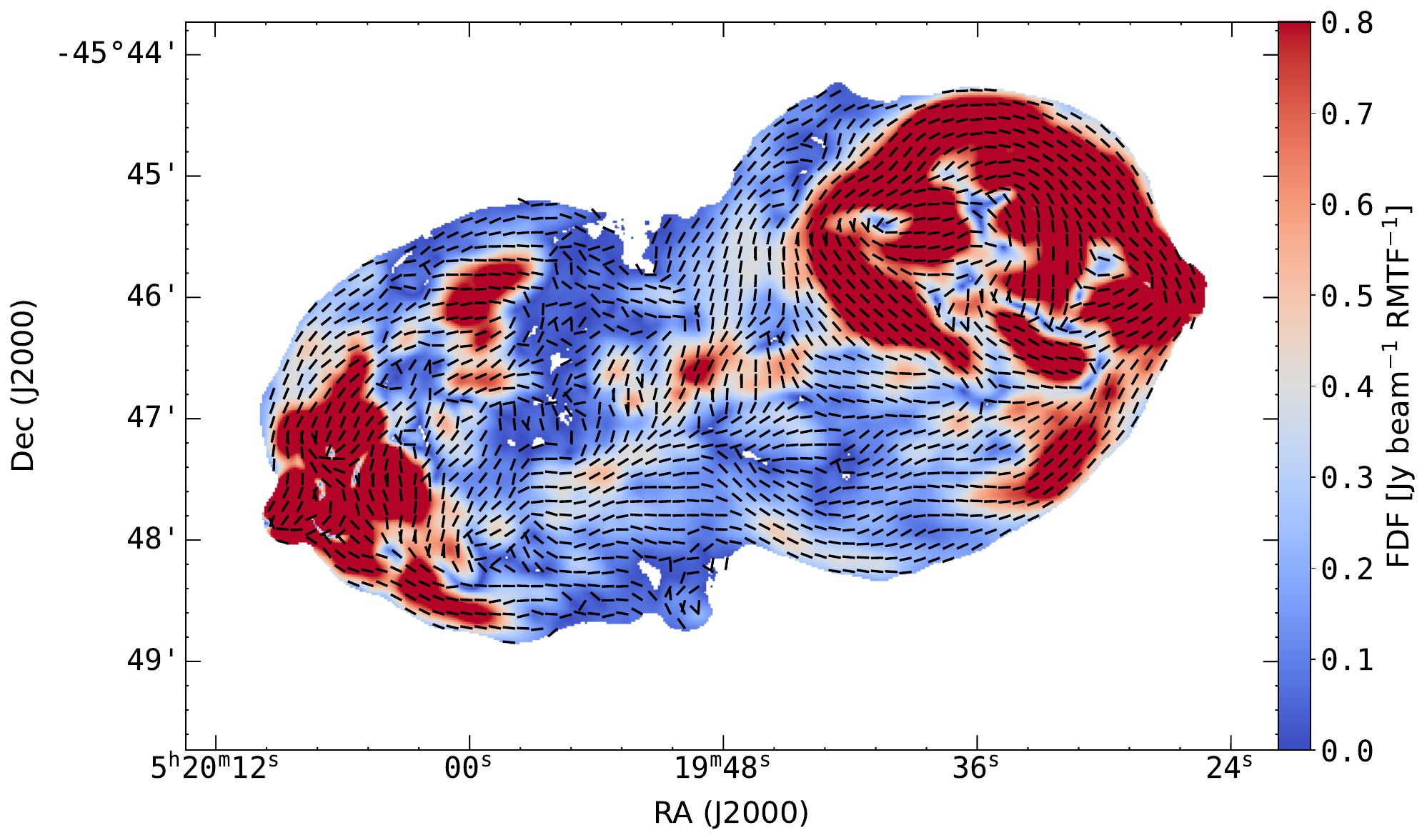}
    \end{minipage}
    \caption{Top panel: The rotation measure map of Pictor A overlaid by Stokes $\mathrm{I}$ radio contours. The RMs across the western lobe are generally smooth, while those across the eastern lobe exhibit more variability. Bottom panel: The polarized intensity map obtained from integrating the cleaned FDF (Eqn. 11 of \citet{Burn1966}), superimposed by the projected magnetic field vectors (polarization vectors rotated by 90$^{\circ}$) whose length is set to 1. The field orientation is well aligned along the lobe edges and the filamentary structures. Both maps are at 11\arcsec by 10\arcsec resolution.} 
    \label{fig:rm-fdf-maps}
\end{figure*}

We performed pixel-by-pixel RM-synthesis across the entirety of Pictor A, resulting in the RM map shown in the top panel of Fig.~\ref{fig:rm-fdf-maps}. Its colour bar ranges between $0$ and $60$ rad m$^{-2}$. Each pixel value in this map represents the Faraday depth corresponding to the highest peak of the cleaned Faraday spectrum. We only show rotation measures in regions of the source whereby $p_{\text{err}} : p > 0.6$, in addition to masking out unphysical values (i.e. where $p \neq [0,1]$). Our lobe edges are the regions where the total intensity falls below 4 mJy beam$^{-1}$.

Generally, the $\mathrm{RM}$ distribution seems relatively similar across the entire source, whereby our derived mean RM for the source was 48 rad m$^{-2}$ with a standard deviation of 10.2 rad m$^{-2}$. The per lobe RM distribution showed that RMs are centred around $49.7$ for the eastern lobe and $46.7$ rad m$^{-2}$ for the western lobe, with widths of $14.4$ and $4.5$ rad m$^{-2}$ respectively. However, patches of significant and sudden RM gradients are observed, especially within the eastern lobe, forming boundary-like structures. A few regions also exhibit negative RMs, such as those near the waist region towards the eastern lobe.

\begin{figure}
    \centering
    \includegraphics[width=1\columnwidth]{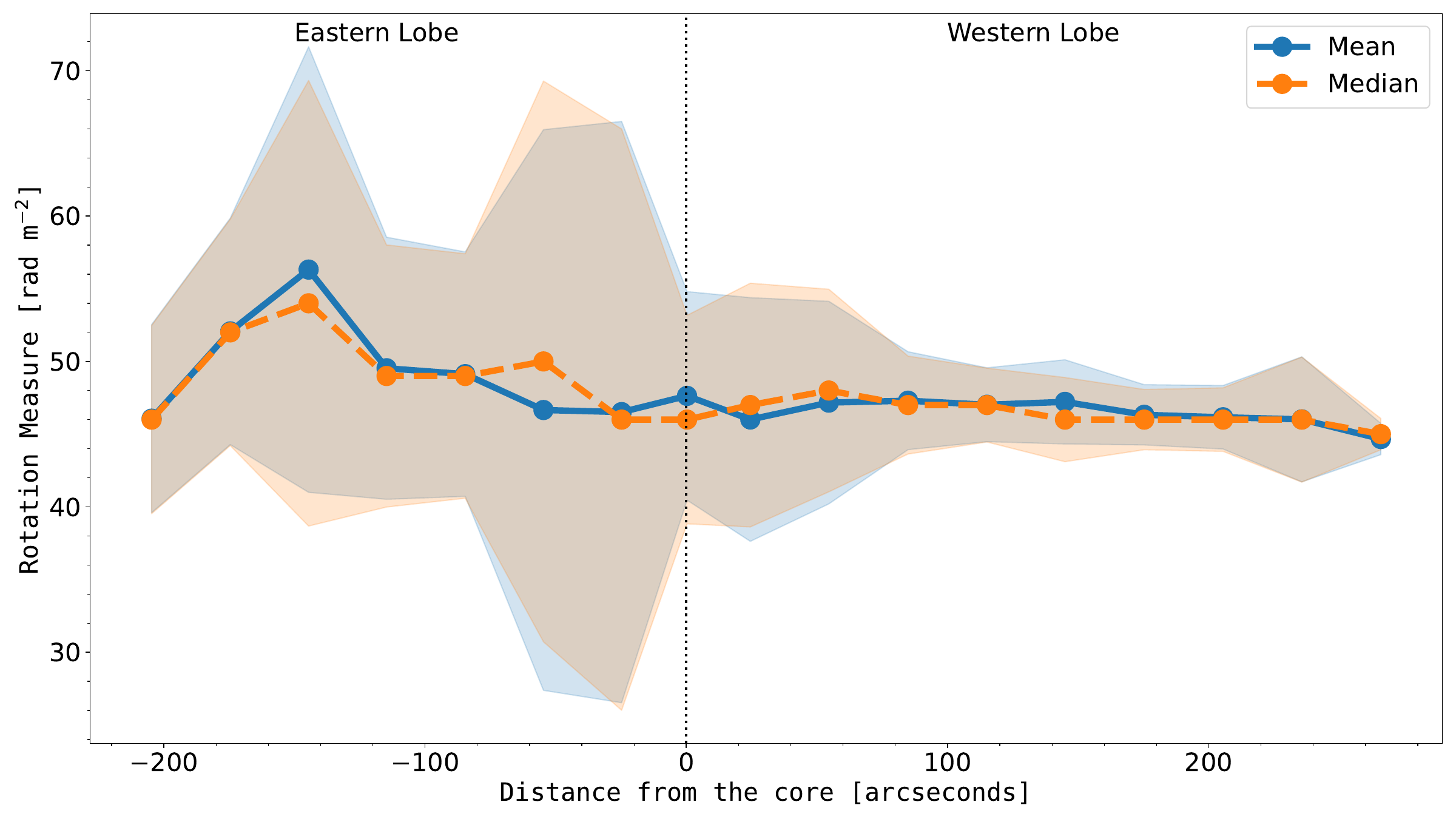}
    \caption{Binned rotation measure profiles with increasing distance from the radio core of Pictor A. Separation of these bins and the distances indicated are similar to those stated for Fig.~\ref{fig:depolarization-profile}. RMs in the western lobe show a decreasing variation with increasing distance from the radio core.}
    \label{fig:rm-lobe-profiles}
\end{figure}

\citeauthor{perley1997} found the average RM to be approximately 43.5$\pm$1.4 rad m$^{-2}$, with the RM for each lobe falling within the margin of error in the L-band. For a more direct comparison, we used our defined lobe boundaries and the VLA L-band RM maps made available to us to extract the per lobe RM distribution from the VLA data. This test yielded RMs for the VLA data centred around 55 and 52 rad m$^{-2}$ and widths of $10.8$ and $6.4$ rad m$^{-2}$ for the eastern and western lobes, respectively. Furthermore, the eastern lobe displays a broader variation than the western lobe in good agreement with the observations of \citeauthor{perley1997}. The difference in our means may be attributed to our high spatial sensitivity. The spread in the RM distribution is not significantly different for each distribution. Simple statistics were also performed based on this RM map to characterise the rotation measure behaviour with distance from the core as illustrated in Fig.~\ref{fig:rm-lobe-profiles}. The variability of the eastern lobe's RMs is more erratic than the western lobe, with regions closer to the core showing more variation.

\citet{haverkorn2015} showed that Galactic contributions within spatially co-located RM largely correlate, making the estimation of contribution on individual lines-of-sight possible. Therefore, we estimated the Galactic contribution towards Pictor A (galactic longitude of $251.6^{\circ}$ and latitude $ -34.63^{\circ}$) by taking the mean and standard deviation of Galactic RM measurements derived by \citet{hutschenreuter2022} (these authors used Bayesian inferencing to cater for noisy data). We found a mean of 23.57 rad m$^{-2}$ and mean standard deviation of $10.87$ rad m$^{-2}$ estimated over a 5$^{\circ}$ region. Our Galaxy may contribute to the mean RM, but it cannot explain the small-scale fluctuations, suggesting that some fraction of the observed RMs may result from the intergalactic medium, X-ray gas or other unknown intervening material.

\subsubsection{Magnetic Fields}
The bottom panel of Fig.~\ref{fig:rm-fdf-maps} shows the orientation of the projected magnetic field (electric field rotated by 90$^\circ$) overlaid on top of the cleaned and integrated Faraday spectra map (obtained by integrating Eqn. 11 of \citet{Burn1966}). We set the length of the field vectors to 1 for demonstration purposes. The magnetic field vectors align along the edges of the lobes. Such alignment is common in radio galaxies \citep[see][]{taylor_vla_1990,Eilek2002,Heald2009,Sebokolodi2020}. Work by \citet{Laing1980} ascribes this effect to the shearing and compression of originally random magnetic fields, making them tangential to the source's boundaries. They argue that the coexistence of aligned magnetic fields and regions of a high fractional polarization only suggests the constriction of the magnetic field to planes along a line-of-sight and not necessarily an originally uniform magnetic field. Moreover, the projected magnetic field vectors also align along the highly linearly polarized filamentary structures within the lobes, especially in the western lobe.

\subsubsection{RM Gradients and Multiple-Peaked Faraday Spectra}
\label{sec:rm-gradients}
In this section, we investigate the cause of the observed RM gradients across Pictor A radio lobes and the presence of Faraday spectra with a single-peak and a wide base, or multiple-peaks.
We find that many LoS with sudden RM changes show multiple high FDF peaks, making it difficult to differentiate from sidelobes. Furthermore, they had much lower polarized emission (the lowest polarized SNR per LoS ranged between 0.4 and 2). The largely flagged out $\lambda^{2}$ data samples could also contribute to this issue because a reduction in the number of $\lambda^{2}$ samples has been shown to increase the RMTF sidelobes \citep[see Fig.7 of][]{Brentjens2005}. A combination of low SNR and large gaps in $\lambda^2$ resulted in higher and erratic side lobes on the deconvolved Faraday spectra (due to sidelobes on the RMTF as illustrated in Fig.~\ref{fig:rmtf}), thus, unstable solutions. 

However, there are regions of high polarized SNR (up to two or three orders of magnitude) showing double/multiple peaks and high RM gradients. These often consist of one dominant peak and a secondary one of lesser intensity, while others show more than two peaks. Spectra showing more than a single peak appear to show "more" complex behaviour in the fractional polarization as a function of wavelength. In most cases, it is difficult to determine whether the LoS depolarizes at longer wavelengths, particularly for those showing repolarization. We will need low-frequency data to properly determine if these lines-of-sight depolarize indefinitely with increasing $\lambda^2$. The ideal type of data would be, for example, similar to that of the wideband polarimetric study of Cygnus A by \citet{Sebokolodi2020}, which consisted of a very broad frequency coverage (2--18 GHz) finely sampled along $\lambda^2$ space. Thus, the authors could characterize the decaying behaviour of fractional polarization for various lines-of-sight (see Figs. 4 and 2 of that work for demonstrations of the various categories of $p(\lambda^2)$ and the RMTF from their data). Their well-sampled $\lambda^2$ space resulted in minimum sidelobes. Using our data, it is impossible to properly characterize the behaviour of $p(\lambda^2)$.

We further investigated whether the multiple peaks were real or whether our high RMTF sidelobe resulted in a false positive peak. To do this, we used the $\mathit{QU}$-fitting method \citep[see for example][]{feain2009,farnsworth2011, osullivan2012, sun_comparison_2015,miyashita2019}, to fit specific models to the data in $\lambda^2$, and determine whether the modelling favoured a model with double peaks or not. Since most of the sight lines across Pictor A exhibited single peaked clean FDF, at least a single Faraday rotating screen was anticipated. Therefore, we explored a single polarized emitting component model coupled with a depolarizing component. For our models, we assumed the external Faraday depolarizing screen (FDS) as the main cause of the observed depolarization. 
The above Faraday effect is modelled using the following:
\begin{equation}
  \label{eqn:efd}
  p = p_{0} e^{-2\sigma^{2}_{RM}\lambda^{4}} e^{2i(\chi_{0} + RM \lambda^{2})},
\end{equation}
where the first exponential term represents depolarization and $\sigma_{\mathrm{RM}}$ is the Faraday dispersion \citep{osullivan2012}. We also explored two emitting components, each with an associated depolarization effect. This is done by summing up the contributions of each component, i.e. summing up Eqn.~\ref{eqn:efd} as $p_{\mathrm{t}} = p_{\mathrm{1}} + p_{\mathrm{2}}$.

Double component models showed reasonably better fits; this was determined using the Bayes' factor method to avoid overfitting. This is not surprising, especially given the structure we see in fractional polarization, which is not expected for a simple uniform foreground screen. A physical interpretation of these results is that multiple Faraday depths or a Faraday thick structure exists along most sight lines across the source. Variations across the observing beam could also lead to the observed behaviour. We note that models with more than two Faraday rotation components were not tested.

\subsubsection{Do the Missing Frequencies Affect RM Measurements?}
\label{sec:missing-freqs}      
One lingering question was whether missing frequency samples affected our RM measurements. As an additional check, we simulated data at 80 evenly spaced channels and set known fiduciary values of fractional polarization, RM, and Stokes $\mathit{I}$ spectral variation. This formed our control dataset, to which Gaussian random noise was added to make it more realistic. We then set the test dataset up to match our actual observed data by excluding the same frequencies that were excised from our channelised images. Specific details about the setup of this experiment are highlighted in Appendix ~\ref{sec:fitting-exp-specs}.

\begin{figure*}
    \centering
    \includegraphics[width=1.94\columnwidth]{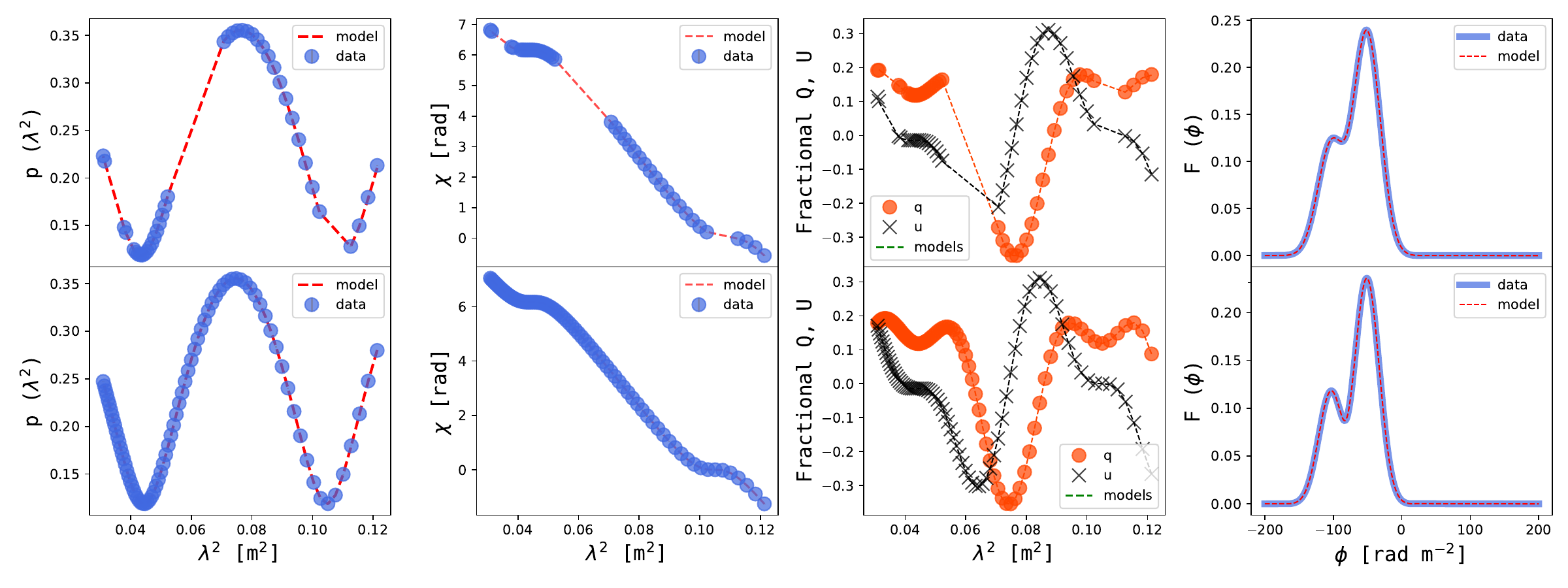}
    \caption{Example plots showing the result of experimental $\mathit{QU}$-fitting on simulated $\mathit{Q}$ and $\mathit{U}$ to ensure that missing frequencies within our data do not affect results. A pair of rows represents a single line-of-sight, with test data on the top panel and the control data on the bottom. Both $\mathit{QU}$-fitting and RM-synthesis could correctly locate the peak amplitude of the emitting components, even with some missing frequency chunks.}
    \label{fig:missing-freqs-ex}
\end{figure*}
Inspection of the results of $\mathit{QU}$-fitting and RM synthesis showed that in both the control and test data cases, it was still possible to recover the correct peak and its corresponding location despite missing $\lambda$ data (some examples are shown in Fig.~\ref{fig:missing-freqs-ex}). However, $\mathit{QU}$-fitting within test cases where the maximum degree of polarization was less than 20 per cent, and two Faraday components had difficulty locating the secondary (lesser) FDF peak position. On the other hand, the strongest FDF peak was correctly identified in almost all cases. Furthermore, the sidelobes were considerably higher, which caused confusion in the validity of a peak.

Two possible interpretations of the multiple peaks observed arise. The first is that there is a single complex continuous Faraday structure (i.e. with a wide Faraday depth) along the line-of-sight of Pictor A. \citet{rudnick2023} (in sec. 4 of their work) have shown that the resolution of the RMTF approximated by \citet{Brentjens2005} incorrectly identifies a continuous tophat Faraday distributions as two narrow-peaked components located at the opposite edges of the original distributions. The second possibility is that two or more Faraday components exist along this LoS. At this point, it is impossible to definitively determine the correct interpretation of the two; wideband data with higher Faraday depth resolution could illuminate this issue better.

\section{Summary and Conclusions}
\label{sec:conclusion}
Using MeerKAT L-band data, we have generated a high-sensitivity map of Pictor A shown in Fig.~\ref{fig:pica}. We observe the bright but unresolved WHS, the double EHS and the radio core. Furthermore, we have confirmed the presence of the kpc scale radio jet from the radio core to the WHS. This radio jet is in good alignment with the X-ray jet from \citet{Hardcastle2016}. However, the counter-jet remains undetected. 

\citet{Hardcastle2016} speculated that if the X-ray emission within the lobes of Pictor A is inverse Compton in nature, it must be co-spatial with synchrotron emission. From the low sensitivity VLA radio observations by \citet{perley1997}, X-ray emission appeared to extend beyond the radio emission, which was strange. However, from our observations, the diffuse emission extending away from the core is now clearly visible (see Fig.~\ref{fig:missing-sync}), with some emission extending beyond the X-ray emission. Thus, we conclude that X-ray emission within the lobes of Pictor A could result from inverse Compton scattering.

Pictor A has also shown spectral variation typical of any FR-II source; it exhibits a flat spectral index in the radio core and hotspots, but spectral steepening occurs within the lobes and especially away from the radio core towards the source edges (see Fig.~\ref{fig:spi-contours} and Table ~\ref{tab:spi-regions} for a summary of the spectral indices). More accurate model fitting and characterisation using wideband data is necessary to distinguish between different variants of spectral ageing models and to determine their parameters.

We have additionally performed a spectropolarimetric study of the lobes of Pictor A (Sec.~\ref{sec:rm-distribution}), where we show that this source exhibits mostly smooth and positive rotation measures within both its lobes, with an average RM of 48.06 rad m$^{-2}$ and a standard deviation of 10.19 rad m$^{-2}$. However, the eastern lobe is associated with a wider RM variation, more sudden RM changes and higher depolarization than the western lobe. We have estimated the average RM in the eastern lobe to be 50 rad m$^{-2}$, while the western lobe averages 47 rad m$^{-2}$. The Galactic contribution was estimated to be $\mathrm{23.57 \pm 16.12}$ rad m$^{-2}$ in the direction of Pictor A. Although Galactic contributions could dominate our observed rotation measures, narrowing down the various contributions with our current data is difficult. This characteristically smooth RM behaviour could indicate a large-scale Faraday rotating screen in the foreground of Pictor A. However, this could also be caused by our data's low Faraday depth resolution directly linked to our limited bandwidth (see Fig.~\ref{fig:rmtf}), which limits our view of smaller Faraday depth scales. 

Furthermore, we have demonstrated that lines-of-sight across Pictor A show varied Faraday spectra, with some showing single, double or multiple peaks. Lines-of-sight with more than one peak indicate the presence of multiple polarized components at different Faraday depths. It is difficult, however, to establish their depolarization characteristics, as the behaviour of fractional polarization is highly variable with wavelength. This directly stems from the limited frequency coverage and missing frequency chunks of our data. Moreover, limited frequency coverage further lowers our depth resolution. Despite this, we have also demonstrated through $\mathit{QU}$-fitting that the missing frequency chunks do not affect our observed rotation measures (Sec.~\ref{sec:missing-freqs}) and that the multiple peaks observed in our data are possibly real, caused by either the existence of multiple Faraday components or the inability of the RMTF to correctly distinguish a single continuous Faraday complex structure (Sec.~\ref{sec:rm-gradients}).  

Therefore, wideband observations will prove helpful in resolving the rotation measure variations and determining whether the observed smoothness across Pictor A is intrinsic to the source or its environs. For example, \citet{Sebokolodi2020} showed that at higher spatial and depth resolution, Cygnus A exhibited higher degrees of polarization and, hence, more polarization structure. Therefore, with better depth and spatial resolutions, the origin of the depolarization and repolarization patches and the RM gradients seen in Figs.~\ref{fig:depolarization} and ~\ref{fig:rm-fdf-maps} could be explained better.

\section*{Acknowledgements}  
We would like to thank M. J. Hardcastle for kindly making the Chandra X-ray images of Pictor A available to us. LALA also thanks U. A. Sob and C. Russeeawon for many helpful consultations and support.

The MeerKAT telescope is operated by the South African Radio Astronomy Observatory, which is a facility of the National Research Foundation, an agency of the Department of Science and Innovation.
OS and LA's research is supported by the National Research Foundation of South Africa (grant number 81737). The financial assistance of the South African Radio Astronomy Observatory (SARAO) towards this research is hereby acknowledged (\url{http://www.sarao.ac.za}).

\section*{Data Availability}
The data are available from the SARAO archive under proposal number SCI-20190418-LS-01. The sub-band Stokes \textit{I}, \textit{Q} and \textit{U} spectral images, and their corresponding MFS images are available at \url{https://doi.org/10.48479/kg33-es24}.

\bibliographystyle{mnras}
\bibliography{pica}

\appendix
\section{Data Simulations and QU-fitting specifics}
\label{sec:fitting-exp-specs}
\begin{figure*}
  \centering
  \includegraphics[width=1.5\columnwidth]{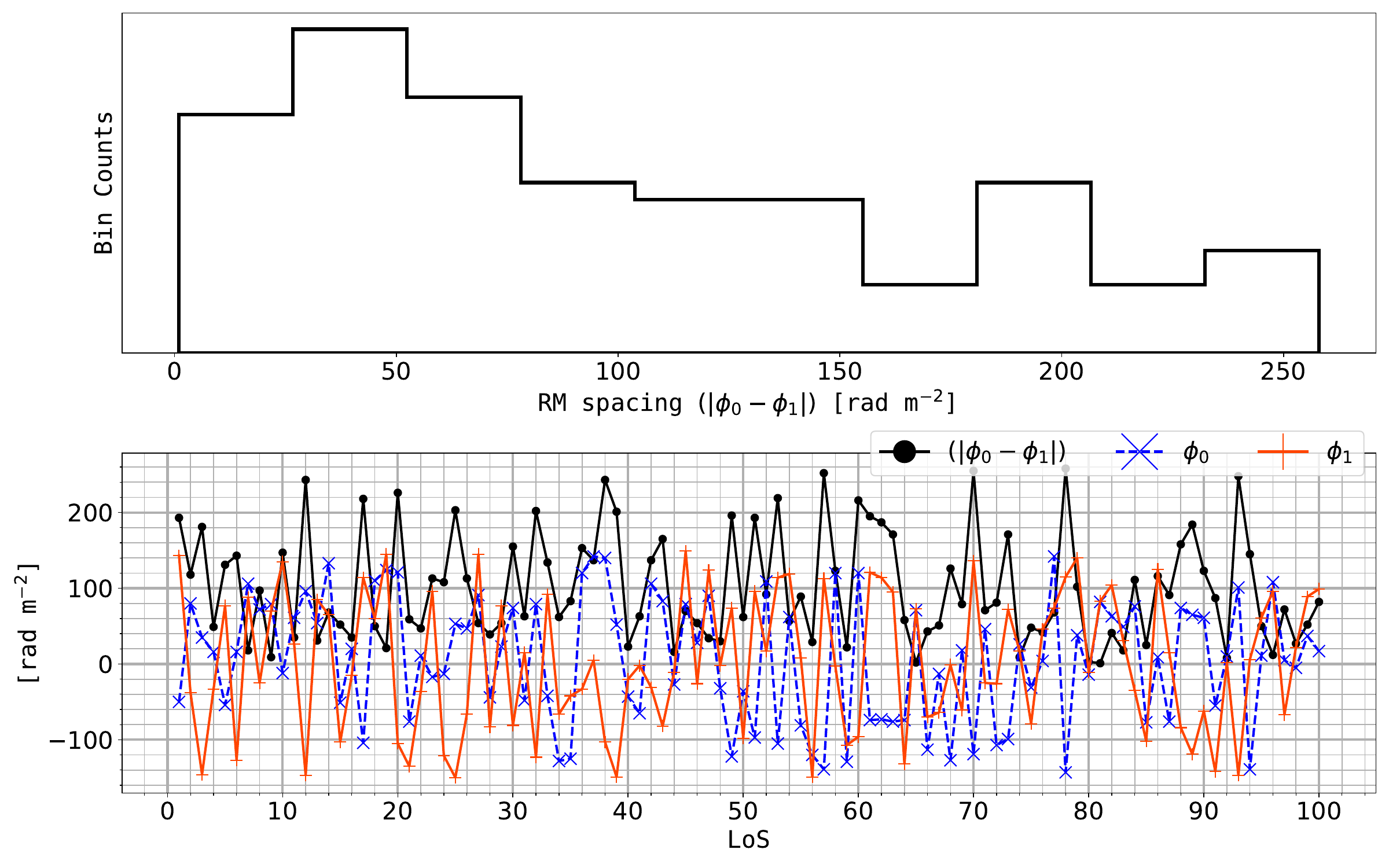}
  \caption{The top panel shows the distribution of RM difference between our two simulated emitting components. The range of our RMs was between $-150$ and 150 rad m$^{-2}$. The bottom panel shows the actual input Faraday depth values; $\phi_0$ is illustrated by the blue dashed line with a cross marker, and $\phi_1$ is illustrated by the red solid line with plus sign markers. The black solid with point markers shows their difference.} 
  \label{fig:exp-rm-spacing}
\end{figure*}

As highlighted in Sec.~\ref{sec:missing-freqs}, data was simulated for 100 lines-of-sight to investigate the effect of missing frequency chunks on our data. We replicated the presence of two Faraday peaks through data simulated using some fiduciary RM and fractional polarization values. The first step was using the 80 frequency channels corresponding to our observation frequency to generate the channelised Stokes $\mathit{I}$ data, with an arbitrary peak of 4 Jy using the following function:
\begin{equation}
    \mathit{I}_{\text{obs}} = \mathit{I}_{\text{true}} (\frac{\nu}{\nu_{0}})^{\alpha},
\end{equation}
where $\alpha$ is the spectral index. A Gaussian random noise was added to make the data more realistic. Using set values of initial fractional polarization, RM and polarization angle, we generated our `observed' fractional polarization data using the model in Eqn.~\ref{eqn:efd}. We also ensured that there is a single dominating peak by setting $p_{0} = 0.5 p_{1}$ (see the bottom panel of Fig.~\ref{fig:exp-rm-spacing}). We used the fractional Stokes relationship: $p = q + iu$ to derive the values of $\mathit{Q}$ and $\mathit{U}$ from their fractional analogues ($\mathit{q, u}$) and our simulated Stokes $\mathit{I}$ data. Furthermore, we included noise values to make the data more realistic. We then stored the fractional polarization, RM and polarization angle for both components and the $\mathit{Q}$ and $\mathit{U}$ data for each frequency channel.

The next step was determining if the peaks and their positions could be recovered accurately, even with missing samples, using $\mathit{QU}$-fitting and RM-synthesis. Therefore, our control experiment involved the data when all available samples were contained. Our test data were similar to the control one, except for the exclusion of channels flagged out in our original data. This was done to replicate our scenario better. This process was repeated 100 times, representing one hundred different lines-of-sight.

\begin{figure*}
  \centering
  \includegraphics[width=1.5\columnwidth]{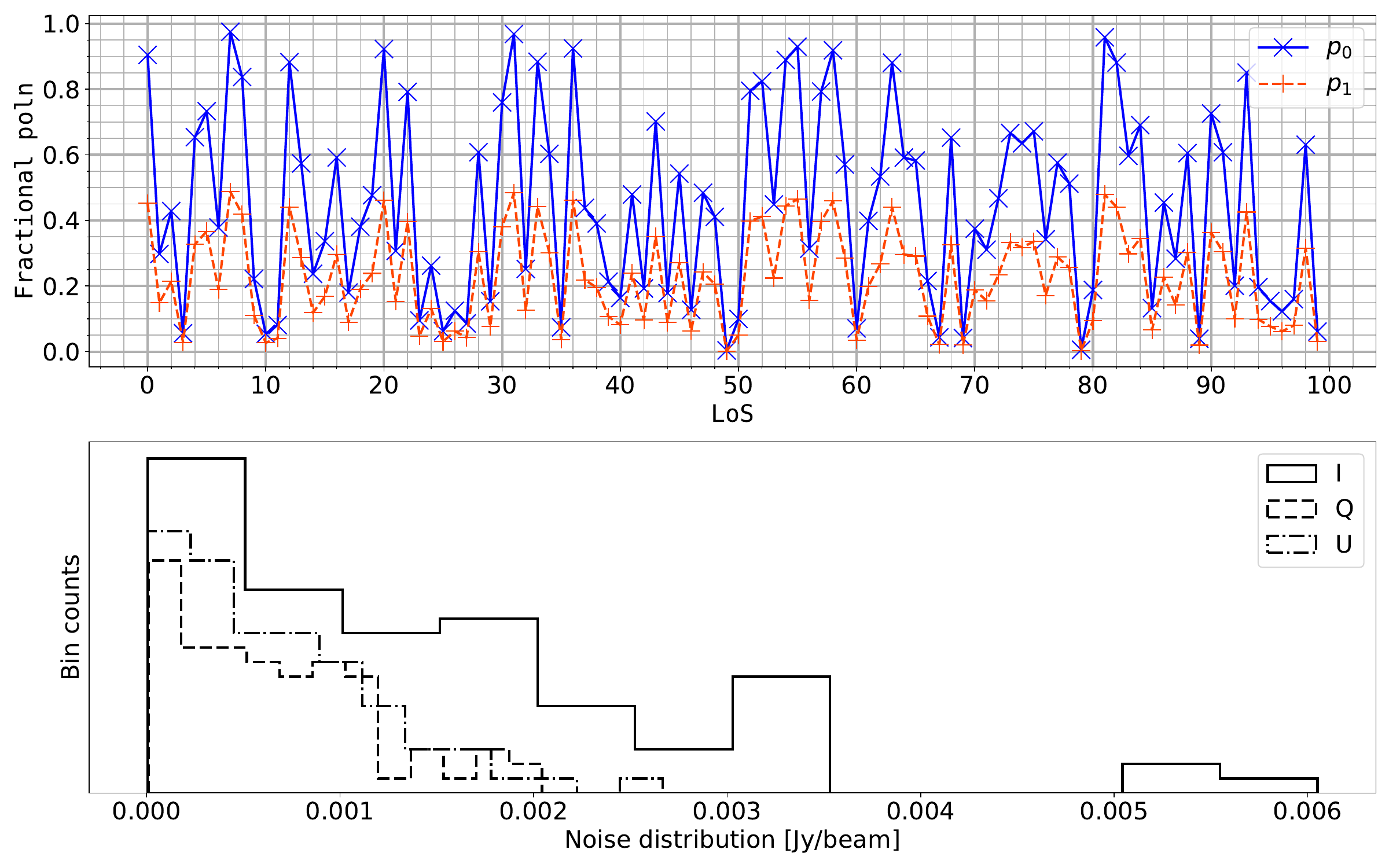}
  \caption{The input fractional polarization values are shown in the top panel. The dominant fractional polarization component is shown in the blue solid line with cross markers, while the secondary component is in the red dashed lines with plus sign markers. This component is always half of the dominant component. \emph{Bottom panel}: Distribution of the input noise added to the simulated \textit{I, Q} and \textit{U} to make it more realistic. All lines-of-sight had the same noise across the band.}
  \label{fig:exp-noise-dist}
\end{figure*}

We chose RM values ranging between $-150$ and $150$ rad m$^{-2}$, as this is the maximum Faraday depth scale our data is sensitive to. These data are shown in Fig.~\ref{fig:exp-rm-spacing}. The polarization angles are between $[-\frac{\pi}{2}, \frac{\pi}{2}]$. The bottom panel of Fig.~\ref{fig:exp-noise-dist} shows the distribution of the artificial noise added to make the $\mathit{I, Q}$, and $\mathit{U}$ data more realistic. The exact values are shown at the top panel of the same figure.
  
\bsp	
\label{lastpage}

\end{document}